\documentclass[prd,preprintnumbers,nofootinbib,floatfix]{revtex4} 
\usepackage{graphicx} 
\usepackage{amsmath}
\usepackage{amsfonts,amsbsy}
\usepackage{amssymb}
\usepackage{appendix}

\def\be{\begin{equation}}
\def\ee{\end{equation}}
\def\bea{\begin{eqnarray}}
\def\eea{\end{eqnarray}}

\def\eq#1{{Eq.~(\ref{#1})}}
\def\fig#1{{Fig.~\ref{#1}}}

\def\beq{\begin{equation}}
\def\eeq{\end{equation}}
\def\bea{\begin{eqnarray}}
\def\eea{\end{eqnarray}}
\def\eq#1{{Eq.~(\ref{#1})}}
\def\fig#1{{Fig.~\ref{#1}}}

\newcommand{\Lb}{\left(}
\newcommand{\Rb}{\right)}

\begin{document}

\title{\bf CGC predictions for p+A collisions at the LHC and signature of QCD saturation }


\author{ Amir H. Rezaeian}
\affiliation{
Departamento de F\'\i sica, Universidad T\'ecnica
Federico Santa Mar\'\i a, Avda. Espa\~na 1680,
Casilla 110-V, Valparaiso, Chile }

\begin{abstract}
We present various predictions for the upcoming p+Pb collisions at $\sqrt{S}=5$ TeV within the color glass condensate (CGC) formalism, including single inclusive charged hadron production, single inclusive prompt photon production, direct photon production, charged hadron multiplicity distribution and photon-hadron azimuthal correlations. Using the running-coupling Balitsky-Kovchegov evolution equation for calculating various observables,  
we show that the main source of uncertainties is due to less constrained initial nuclear saturation scale. This gives rise to rather large theoretical uncertainties for nuclear modification factor $R_{pA}$ at the LHC. Nevertheless, we propose a simple scheme in which one can still test the main dynamics of the CGC/saturation in p+A collisions at the LHC.

\end{abstract}

\maketitle

\section{Introduction}

The upcoming proton-lead (p+Pb) collisions at the LHC will provide crucial  benchmarks for understanding the characteristic of  Quark Gluon Plasma produced in heavy ion collisions at the LHC and RHIC. Moreover, p+A collisions run has its own merits, namely it can be vital testing grounds for novel nontrivial QCD dynamics which otherwise it cannot be unambiguously  explored. It is generally believed that a system of partons (gluons) at high energy (or small Bjorken-x) forms a new state of matter where the gluon distribution saturates and non-linear coherence phenomena dominate \cite{sg}. Such a system is endowed with a new dynamical momentum scale, the so-called saturation scale which controls the main characteristic of the particle production. The color glass condensate (CGC) approach was proposed as a framework to study small-x and saturation physics \cite{mv}. The CGC formalism is an effective perturbative (weak-coupling) QCD theory  in which one systematically re-sums quantum corrections which are enhanced by large logarithms of 1/x and also incorporates high gluon density effects which are important at small x and for large nuclei, for a review see e.g. Ref.\,\cite{cgc-review1}.   

The CGC formalism has been successfully applied to many processes in high energy collisions. Examples are structure functions (inclusive and diffractive) in Deeply Inelastic Scattering of electrons on protons or nuclei, and particle production in proton-proton, proton-nucleus and nucleus-nucleus collisions at RHIC, see Ref.\,\cite{cgc-review1} and references therein. The observed suppression of the single inclusive hadron production in deuteron-nucleus collisions (d+A) at forward rapidities at RHIC \cite{exp} is perhaps among the most spectacular evidence of the QCD gluon saturation. There are, however, alternative phenomenological approaches which describe the same data.  Therefore, the upcoming p+A run at the LHC which covers a much wider kinematic region, can provide a unique opportunity to understand the underlying dynamics of particle production at small x.

The key ingredient of the particle production at leading logarithmic accuracy which captures the main saturation dynamics is the universal color dipole cross-section, the imaginary part of the quark-antiquark scatterings amplitude.  It is universal, in a sense that for different processes, the cross-section can be written in terms of the same object, see e.g. Refs.\,\cite{u1,u2}. It has been recently proven that in the large-$N_c$ limit, all multi-particle production processes in the collision of a dilute system off a dense (e.g, p+A collisions) can, up to all orders in the strong-coupling, be described in terms of only dipoles and quadrupoles (quadrupoles can be also approximated to dipoles) \cite{u2}.  The color-dipole amplitude satisfies the Jalilian-Marian-Iancu-McLerran-Weigert-Leonidov-Kovner (JIMWLK) evolution equations \cite{jimwlk}. 
In the large $N_c$ limit, the coupled JIMWLK equations are simplified to the Balitsky-Kovchegov (BK) equation \cite{bk,bb}, a closed-form equation for the rapidity evolution of the dipole amplitude. Unfortunately, at the moment, we do not know how to deal with impact-parameter dependence of the BK (or JIMWLK) evolution equation \cite{bk-c,bk-b}, and impact-parameter dependence of the dipole amplitude is generally ignored.  Another current drawback is that numerical solutions of the full next-to-leading logarithmic expressions \cite{nlo} are not yet available, only running coupling corrections to the leading log kernel have been considered in phenomenological applications, the so-called running-coupling BK equation \cite{rcbk}.  Moreover, the BK evolution only provides the rapidity/energy evolution of the dipole,  the initial profile of the dipole  amplitude and its parameters still need to be modeled and constrained by experimental data. Unfortunately, the current world-wide small-x data are very limited, and cannot uniquely fix the initial parameters of the dipole  amplitude \cite{jav1}. This problem is more severe for determining dipole scattering amplitude on nuclear target. This leads to rather large unavoidable theoretical uncertainties for various CGC predictions in p+A collisions at the LHC. Here, we propose a simple scheme in which one can overcome this problem and still test the model at the LHC.  

In this letter, we provide a compilation of various predictions for p+Pb collisions at  $\sqrt{S}=5$ TeV  at different rapidities based on the CGC approach. This includes the total charged hadron multiplicity distribution at different centralities, nuclear modification factor for single inclusive charged hadron production, nuclear modification factor for single inclusive prompt photon production (and direct photon production), and azimuthal photon-hadron correlations.  These observables have been already studied in the CGC framework but at different energies at the LHC, namely $\sqrt{S}=4.4$ and $8.8$ TeV \cite{me-pa,ja1,ja2,me-ph}. Here we also extend the previous studies by quantifying theoretical uncertainties coming from 
our freedom to choose different values for the factorization scale $Q$, the strong-coupling $\alpha_s$, and initial saturation scale of proton and nucleus. 

\section{Theoretical framework and main formalism}

\subsection{Single inclusive hadron production in p+A collisions; $k_T$ factorization, and hybrid formalism } 
In the CGC approach, the gluon jet production in p+A collisions can be described by $k_T$-factorization \cite{kt},
\begin{equation} \label{M1}
\frac{d \sigma}{d y \,d^2 p_{T}}=\frac{2\alpha_s}{C_F}\frac{1}{p^2_T}\int d^2 \vec k_{T} \phi^{G}_{p}\Lb x_1;\vec{k}_T\Rb \phi^{G}_{A}\Lb x_2;\vec{p}_T -\vec{k}_T\Rb,
\end{equation}
where $C_F=(N_c^2-1)/2N_c$ with $N_c$ being the number of colors,
$x_{1,2}=(p_T/\sqrt{S})e^{\pm y}$, $p_T$ and $y$ are the
transverse-momentum and rapidity of the produced gluon
jet with $\sqrt{S}$ being the nucleon-nucleon center of mass energy. $\phi^{G}_{A}(x_i;\vec k_T)$ denotes the unintegrated gluon
density and is the probability to find a gluon that carries $x_i$
fraction of energy with $k_T$ transverse momentum in the projectile (or target) A. The unintegrated gluon
density is related to the color dipole forward scattering amplitude,  
\beq \label{M2}
\phi^G_A\Lb x_i;\vec{k}_T\Rb=\frac{1}{\alpha_s} \frac{C_F}{(2 \pi)^3}\int d^2 \vec b_T\,d^2 \vec r_T
e^{i \vec{k}_T\cdot \vec{r}_T}\nabla^2_T \mathcal{N}_A\Lb x_i; r_T; b_T\Rb,
\eeq
with notation
\beq \label{M3}
\mathcal{N}_A\Lb x_i; r_T; b_T \Rb =2 \mathcal{N}_F\Lb x_i; r_T; b_T \Rb - \mathcal{N}^2_F\Lb x_i; r_T; b_T \Rb,
\eeq
where $r_T$ denotes the dipole transverse size and $b_T$ is the
impact parameter of the scattering. Throughout this paper, the subscript $T$ stands for the transverse component. For the value of strong-coupling
$\alpha_s$ we employ the running coupling prescription used in
Refs.~\cite{me-pp,me-aa1,me-pp2,me-aa}. Namely, in \eq{M1} we replace $\alpha_s$ by $\alpha_s(p_T)$, and in \eq{M2} we replace $\alpha_s$ by  $\alpha_s(Q_s(x_i))$ with $Q_s(x_i)$ being the saturation scale in the projectile or target \cite{me-pp}. 
The most important ingredient of the $k_T$ factorization which encodes the saturation dynamics is the
fundamental (or adjoint) dipole amplitude, the imaginary part of
the forward quark anti-quark scattering amplitude on a proton or
nucleus target $N_{p,A}\Lb x_i; r_T; b \Rb$.  The dipole cross-section can be
computed via the perturbative nonlinear small-x Balitsky-Kovchegov (BK) quantum evolution equation \cite{bk}, see below.  
Note that the $k_T$ factorization has been proven at the leading log approximation for scatterings of a dilute system on a dense one (such as proton-nucleus collisions) and includes BFKL type gluon emissions with gluon fusion effects between the projectile and target, and also gluon radiations from the produced gluons \cite{kt}. This formulation has been independently verified by many authors, see e.g. Ref.\,\cite{kt-rest}. 
 
The rapidity distribution of the inclusive mini-jet production can be calculated from \eq{M1}, 
\beq \label{PO1}
\frac{d N_{\mbox{jet}}}{d \eta}\,\,=\,\,\frac{K}{\sigma_{s}}\int d^2 p_T \,h[\eta, p_T, m_{jet}] \frac{d \sigma}{d y \,d^2 p_{T}}\left[\eq{M1}\right],
\eeq
where $\eta$ is the pseudorapidity and $h[\eta,p_T, m_{jet}]$ is the Jacobian
which takes account of the difference between rapidity $y$ and the measured pseudorapidity $\eta$ \cite{me-pp}.  In the above, we introduced a $K$-factor to effectively incorporate the missing higher-order corrections.  $\sigma_{s}$ is the average interaction area which depends on the kinematics and  is related to the average impact parameter of the inclusive production of the mini-jet \cite{me-pp}. By introducing mini-jet mass $m_{jet}$ which mimics the pre-hadronization effect, one can regularize the infrared divergences of the $k_T$ factorization cross-section \cite{me-pp,me-aa1,me-pp2,me-pa,me-aa}.  Finally, in order to relate the mini-jet yield to the produced hadron, we employ the so-called Local Parton-Hadron Duality principle \cite{lphd} assuming that the form of the rapidity distribution will not be distorted by the jet decay and only
a numerical factor will differ the mini-jet spectrum from the hadron one. Note that the main contribution of the integrand in \eq{PO1} comes from low transverse momentum (typically less than $1\div 3$ GeV), therefore using fragmentation function is less justifiable.

In the $k_T$-factorized approach,  partons in both 
the projectile and target are assumed to be at very small $x$ in order the CGC
formalism and small-x resummation to be applicable to both the projectile and target. This
approach is valid as long as one stays away from the projectile
fragmentation region. However, for the projectile fragmentation region at forward collisions, an alternative approach was developed
in~\cite{dhj,inel}, the so-called hybrid approach, which is better suited where one treats the projectile wave-function
perturbatively within the standard collinear factorization approach using the standard DGLAP picture while treating
the target by CGC methods. The cross section for single inclusive hadron production at leading twist approximation, in asymmetric collisions such as p+A  ones, in the CGC formalism is given by~\cite{dhj,inel},
\begin{eqnarray}\label{final}
\frac{dN^{p A \rightarrow h X}}{d^2p_T d\eta}&=&\frac{K}{(2\pi)^2}\Bigg[\int_{x_F}^1 \frac{dz}{z^2} \Big[x_1f_g(x_1,Q^2)N_A(x_2,\frac{p_T}{z})D_{h/g}(z,Q)+\Sigma_qx_1f_q(x_1,Q^2)N_F(x_2,\frac{p_T}{z})D_{h/q}(z,Q)\Big]\nonumber\\
&+& \frac{\alpha_s^{in} }{ 2\pi^2} \int_{x_F}^1 \frac{dz}{z^2}\frac{z^4}{p_T^4}\int_{k_T^2<Q^2}d^2k_T k_T^2 N_F(k_T,x_2)\int_{x_1}^1\frac{d\xi}{\xi}\Sigma_{i,j=q,\bar q, g}w_{i/j}(\xi)P_{i/j}(\xi)x_1f_j(\frac{x_1}{\xi}, Q)D_{h/i}(z,Q)\Bigg],
\end{eqnarray}
where $\alpha_s^{in} $ behind the inelastic term is the QCD strong-coupling and denoted by subscript "$in$" in order to be different with the strong-coupling $\alpha_s$-running in the rcBK equation. Note that in the hybrid formulation given above, in principle the strong coupling in the projectile which is in dilute regime can be different from the one in the rcBK description of dense target. The scale of $\alpha_s^{in} $ cannot be determined at its current approximation, and for that a full calculation upto NNLO is required.  We will later consider the implication of various values for $\alpha_s^{in}$. $f_j(x,Q^2)$ is the parton distribution function (PDF) of the
incoming proton which depends on the light-cone momentum fractions $x$
and the hard scale $Q$. The function $D_{h/i}(z,Q)$ is the
hadron fragmentation function (FF) of $i$`th parton to the final
hadron $h$ with a momentum fraction $z$.  The indices $q$ and $g$ denote quarks and gluon, with a summation over different flavors being implicit. The inelastic weight function $w_{i/j}$ and the DGLAP splitting functions $P_{i/j}$ are given in Ref.\,\cite{inel}. The longitudinal momentum fractions $x_1$ and $x_2$ are defined as follows,
\begin{equation}\label{xs}
x_F\approx \frac{p_T}{\sqrt{S}}e^{\eta}; \ \ \ \ x_1=\frac{x_F}{ z}; \ \ \ \ \ x_2=x_1e^{-2\eta}. 
\end{equation}
Note that in the hybrid formalism \eq{final}, in contrast to the $k_T$ factorization \eq{M1}, hadronization can be treated similar to the standard pQCD  due to  resummation of collinear singularities via DGLAP evolution for the incoming and outgoing parton.  The expersion given in \eq{final}  can be understood in a simple intuitive picture.  The first two terms correspond to elastic contribution, namely an incoming parton scatters elastically with the CGC target. This incoming parton with initial zero transverse momentum picks up transverse momentum of order saturation scale after multiple scatterings on the dense target. The last term in \eq{final}  gives the inelastic contribution to the inclusive hadron production. Namely, the projectile parton can also interact with target inelastically with small transfer momentum exchanges. This decoheres the pre-existing high-$p_T$ parton from the hadron wave function and releases it as an on-shell particle. These high-$p_T$ partons in the projectile wave function arise due to DGLAP splitting of partons.  

In above, the amplitude $N_F$ ($N_A$) is the two-dimensional Fourier
transformed of the imaginary part of the forward dipole-target
scattering amplitude $\mathcal{N}_{A(F)}$ in the fundamental (F) or adjoint (A)
representation,
\begin{equation} \label{ff}
N_{A(F)}(x,k_T)=\int d^2\vec r e^{-i\vec k_T.\vec r}\left(1-\mathcal{N}_{A(F)}(r,Y=\ln(x_0/x))\right),
\end{equation}
where $r=|\vec r|$ is the dipole transverse size. The dipole forward scattering amplitude incorporates small-x dynamics and can be calculated via the running-coupling BK (rcBK) evolution equation. The rcBK equation has the following simple form:
\begin{equation}
  \frac{\partial\mathcal{N}_{A(F)}(r,x)}{\partial\ln(x_0/x)}=\int d^2{\vec r_1}\
  K^{{\rm run}}({\vec r},{\vec r_1},{\vec r_2})
  \left[\mathcal{N}_{A(F)}(r_1,x)+\mathcal{N}_{A(F)}(r_2,x)
-\mathcal{N}_{A(F)}(r,x)-\mathcal{N}_{A(F)}(r_1,x)\,\mathcal{N}_{A(F)}(r_2,x)\right],
\label{bk1}
\end{equation}
with $\vec r_2 \equiv \vec r-\vec r_1$. The only external input for the rcBK non-linear equation is the initial condition for the evolution which is taken to have the following form motivated by McLerran-Venugopalan (MV) model \cite{mv},  
  \begin{equation}
\mathcal{N}(r,Y\!=\!0)=
1-\exp\left[-\frac{\left(r^2\,Q_{0s}^2\right)^{\gamma}}{4}\,
  \ln\left(\frac{1}{\Lambda\,r}+e\right)\right],
\label{mv}
\end{equation}
where the onset of small-x evolution is assumed to be at
$x_0=0.01$, and the infrared scale is taken $\Lambda=0.241$ GeV \cite{jav1}. 
The only free parameters in the above are $\gamma$ and the initial saturation scale $Q_{0s}$, with a notation $s=p$ and, $A$ for a proton and nuclear target, respectively. We will later consider uncertainties coming from our freedom to choose among different parameter sets of the rcBK description of the proton and nucleus.

\subsection{ Direct photon production and photon-hadron correlation in p+A collisions} 

The cross section for semi-inclusive prompt photon-quark production in p+A collisions at the leading twist approximation in the CGC formalism is given by  \cite{pho-cgc1,pho-cgc2},   
\bea
&&{d\sigma^{q\, A \rightarrow q(l)\,\gamma(p^\gamma)\, X}
\over d^2\vec{b_T}\, d^2\vec{p_T}^{\gamma}\, d^2\vec{l_T}\, d\eta_{\gamma}\, d\eta_h} =
{Ke_q^2\, \alpha_{em} \over \sqrt{2}(4\pi^4)} \, 
{p^-\over  (p_T^\gamma)^ 2 \sqrt{S}} \,
{1 + ({l^-\over k^-})^2 \over
[p^- \, \vec{l_T} - l^- \vec{p_T}^\gamma]^2}\nonumber \\
&&\delta [x_q - {l_T \over \sqrt{S}} e^{\eta_h} - {p_T^\gamma \over \sqrt{S}} e^{\eta_\gamma} ] \,
\bigg[ 2 l^- p^-\, \vec{l_T} \cdot \vec{p_T}^\gamma + p^- (k^- -p^-)\, l_T^2 + l^- (k^- -l^-)\, (p_T^\gamma)^2 \bigg] N_F (|\vec{l_T} + \vec{p_T}^\gamma|,  x_g) ,
\label{cs}
\eea
where $p^\gamma$, $l$,  and $k$  are $4$-momenta of the produced prompt photon, outgoing quark and  projectile quark, respectively. 
Again, a $K$-factor was introduced to absorb higher-order corrections. 
The light-cone fraction $x_q$ is the ratio of the incoming quark to proton energies, namely $x_q =k^-/\sqrt{S/2}$. The pseudorapidities of 
outgoing prompt photon $\eta_\gamma$ and quark $\eta_h$ are defined via $p^-={p_T^\gamma \over \sqrt{2}} e^{\eta_{\gamma}}$
and $l^-={l_T \over \sqrt{2}} e^{\eta_h}$. The angle between the final-state quark and prompt photon is denoted by $\Delta \phi$ and defined via $\cos(\Delta \phi) \equiv {\vec{l}_T \cdot \vec{p}_T^\gamma \over  l_t p_T^\gamma}$. We only consider here light hadron production, therefore at high transverse momentum, the rapidity and pseudorapidity are the same. 
In \eq{cs}, $N_F (p_T, x_g)$ is again Fourier transformed of the dipole amplitude which satisfies the rcBK equation (\ref{bk1}). 
The semi-inclusive photon-hadron production in proton-nucleus collisions can be obtained from partonic cross-section \eq{cs}  by convolution of quark and 
antiquark distribution functions of a proton and the quark-hadron fragmentation function,
\begin{eqnarray}\label{qh-f}
\frac{d\sigma^{p\, A \rightarrow h (p^h)\, \gamma (p^\gamma)\, X}}{d^2\vec{b_T} \, d^2\vec{p_T}^\gamma\, d^2\vec{p_T}^h \,  
d\eta_{\gamma}\, d\eta_{h}}&=& \int^1_{z_{f}^{min}} \frac{dz_f}{z_f^2} \, 
 \int\, dx_q\,
f _q(x_q,Q^2)  \frac{d\sigma^{q\, A \rightarrow q(l)\,\gamma(p^\gamma)\, X}}
{ d^2\vec{b_T}\, d^2\vec{p_T}^{\gamma}\, d^2\vec{l_T}\, d\eta_{\gamma}\, d\eta_h}  D_{h/q}(z_f,Q^2),
\end{eqnarray}
where $p^h_T$ is the transverse momentum of the produced hadron. A summation over the 
quark and antiquark flavors in the above expression should be understood. The light-cone momentum fraction $x_q, x_{\bar q}, x_g$ in Eqs.\,(\ref{cs},\ref{qh-f}) are related to the transverse momenta and
rapidities of the produced hadron and prompt photon via \cite{ja2},
\begin{eqnarray}\label{qh-k}
x_q&=&x_{\bar{q}}=\frac{1}{\sqrt{S}}\left(p_T^\gamma\, e^{\eta_{\gamma}}+\frac{p_T^h}{z_f}\, e^{\eta_h}\right),\nonumber\\
x_g&=&\frac{1}{\sqrt{S}}\left(p_T^\gamma\, e^{-\eta_{\gamma}}+ \frac{p_T^h}{z_f}\, e^{-\eta_{h}}\right),\nonumber\\
z_f&=&p_T^h/l_T, \hspace{1 cm} \text{with}~~~~~ z_{f}^{min}=\frac{p_T^h}{\sqrt{S}}
\left(\frac{e^{\eta_h}}
{1 - {p_T^\gamma\over \sqrt{S}}\, e^{\eta_{\gamma}}}\, 
\right).\label{z_f}\label{ki1}\
\end{eqnarray}

The single inclusive prompt photon cross section in the CGC framework can be obtained from
\eq{cs} by integrating over the momenta of the final state quark. The cross-section of single inclusive prompt photon production can be divided into two contributions of fragmentation and direct photon \cite{ja2}: 
\begin{eqnarray}\label{pho2}
\frac{d\sigma^{q\, A \rightarrow \gamma (p^\gamma) \, X}}{d^2 \vec{b_T} d^2 \vec{p_T}^\gamma d\eta_{\gamma}}&=&
\frac{d\sigma^{\text{Fragmentation}}}{d^2 \vec{b_T} d^2 \vec{p_T}^\gamma d\eta_{\gamma}}+\frac{d\sigma^{\text{Direct}}}{d^2 \vec{b_T} d^2 \vec{p_T}^\gamma d\eta_{\gamma}}, 
\\
&=&\frac{K}{(2\pi)^2}\Big[\frac{1}{z}\, D_{\gamma/q}(z, Q^2)\, 
N_F(x_g,p_T^\gamma/z) + 
 \frac{e_q^2 \alpha_{em}}{2\pi^2}z^2[1+(1 - z)^2]\frac{1}{(p_T^\gamma)^4}
\int_{l_T^2<Q^2}d^2\vec{l_T}\,l_T^2\, 
N_F(\bar{x}_g,l_T)\Big], \nonumber\
\end{eqnarray}
where $D_{\gamma/q}(z,Q^2)$ is the leading order quark-photon fragmentation function \cite{own}. Similar to the hybrid formalism for the inclusive hadron production \eq{final}, $Q$ is a hard-scale. One should note that above expersion was obtained using a hard cutoff to subtract the collinear singularity \cite{ja2}. This may result in a mismatch between the finite corrections to our results and those that are included in parameterizations of photon fragmentation function. However, this mismatch is a higher order effect in the coupling constant and its proper treatment requires a full NLO calculation which is beyond the scope of this letter. In order to somehow quantify possible errors associated with the approximation made in \eq{pho2}, we will consider different photon fragmentation functions, and also different values for the hard-scale $Q$. 
In order to relate the partonic cross-section given by \eq{pho2} to prompt photon production in p+A collisions, we convolute \eq{pho2} with quark and antiquark distribution functions of the projectile proton, 
 \begin{equation}\label{pho4}
\frac{d\sigma^{p\, A \rightarrow \gamma (p^\gamma) \, X}}{d^2\vec{b_T} d^2\vec{p_T}^\gamma d\eta_{\gamma}}=  
\int_{x_q^{min}}^1 d x_q f_q(x_q, Q^2)
\frac{d\sigma^{q (q^h) \, A \rightarrow \gamma (p^\gamma) \, X}}{d^2 \vec{b_T} d^2\vec{p_T}^\gamma d\eta_{\gamma}},
\end{equation}
where a summation over different quarks (antiquarks) flavors is implicit. The light-cone fraction 
variables $x_g,\bar{x}_g,z$ in Eq.~(\ref{pho2},\ref{pho4})  are related to the transverse momentum of the produced prompt photon and its rapidity \cite{ja2}, 
\begin{eqnarray}\label{pho5}
x_g&=& x_q \, e^{-2\, \eta_\gamma}, \nonumber\\
\bar{x}_g &=& \frac{1}{x_q\, S} \left[{(p_T^\gamma)^2\over z} + \frac{(l_T-p_T^\gamma)^2}{1-z}\right], \nonumber\\
z&=& \frac{p_T^\gamma}{x_q\, \sqrt{S}}e^{\eta_{\gamma}} , 
\hspace{1 cm} \text{with}~~~~~ x_q^{min}=\frac{p_T^\gamma}{\sqrt{S}}e^{\eta_{\gamma}}. \
\end{eqnarray}
Similar to  the inclusive hadron production, the main ingredient of photon-hadron and prompt photon production cross-section is the universal dipole amplitude. One should note that the light-cone fraction variables  which enter in the cross-section for the inclusive photon \eq{pho5} and semi-inclusive photon-hadron \eq{qh-k}, and single inclusive hadron production \eq{xs} are different.

\section{Discussion and Predictions }

\begin{figure}[t]                                                                                          
                               \includegraphics[width=8 cm] {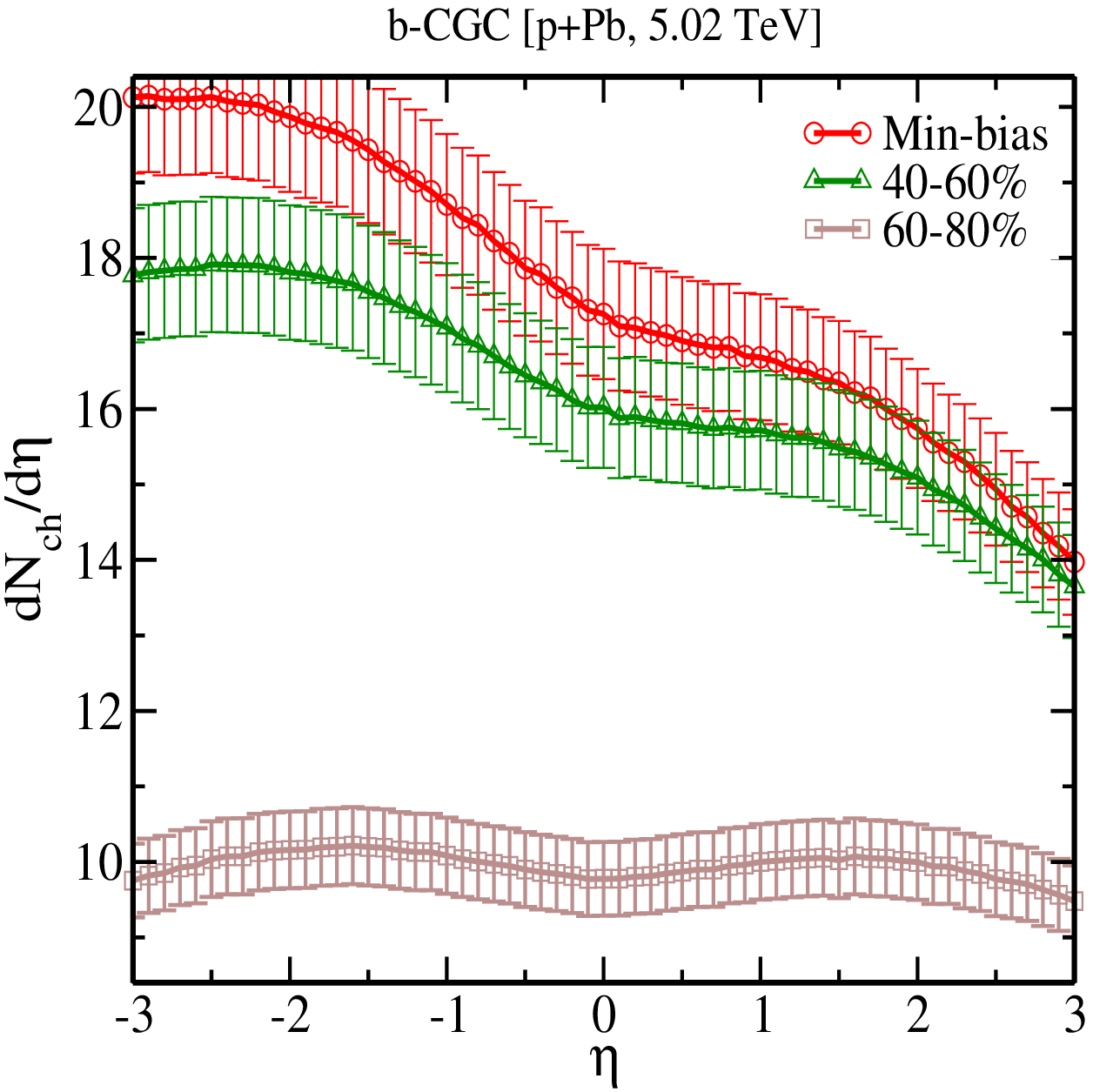}      
 \includegraphics[width=8 cm] {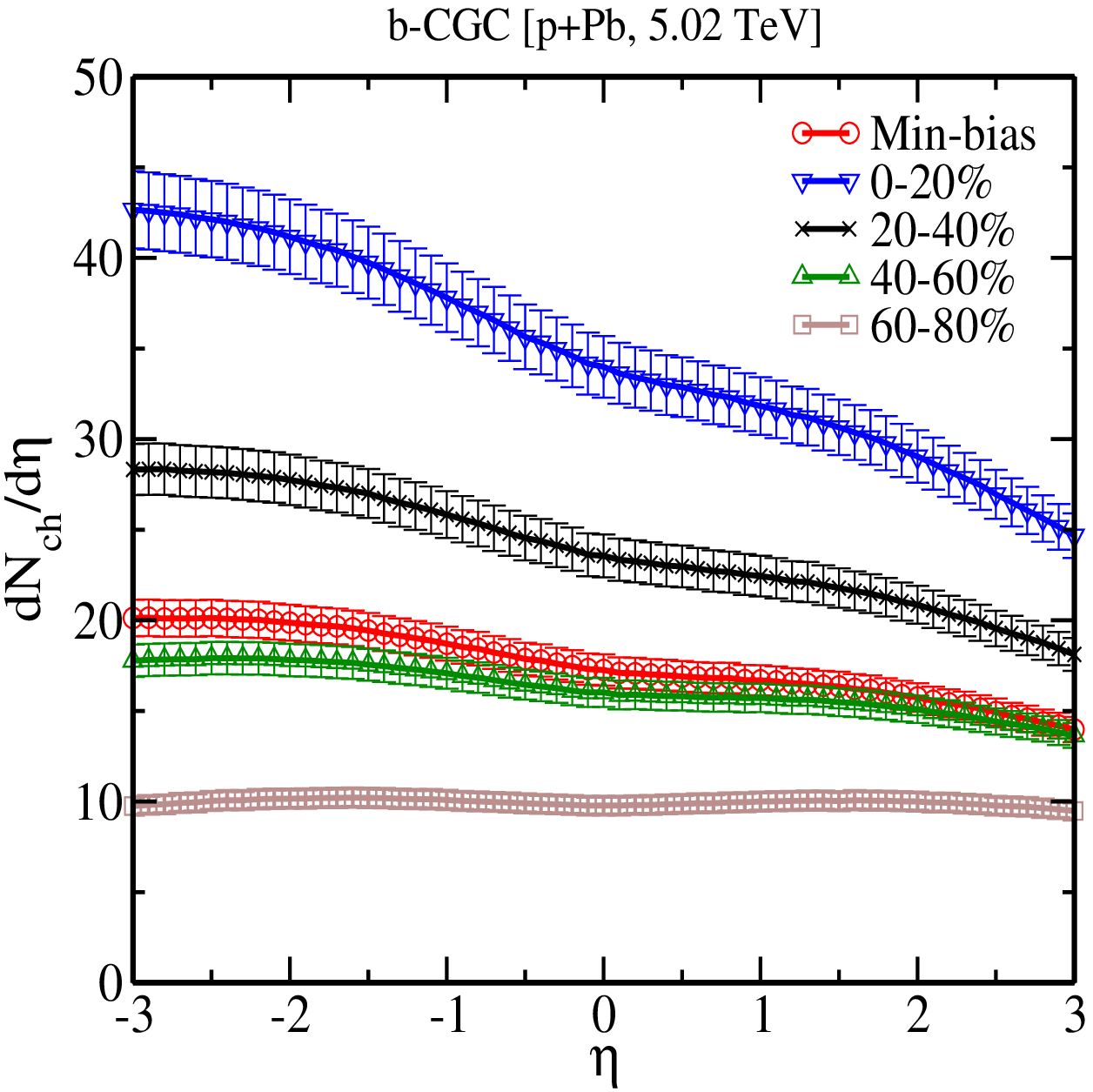}                             
\caption{Pseudorapidity distribution of the
                                  charged particles production in  p+Pb 
                                   collisions at the LHC
                                  $\sqrt{S}=5.02$ TeV at various centralities, in right panel, from top to down: $0-20\%$, $20-40\%$, minimum-bias, $40-60\%$, $60-80\%$. 
 Left panel: for a better comparison, we show again theoretical curves for more peripheral centrality bins (the theoretical curves in both panels are the same).  
The theoretical uncertainties of about $5\%$ due to fixing the over-all normalization at RHIC are also shown.}
\label{fig-m}
\end{figure}

We first present our results for the total charged hadron multiplicity distribution in p+Pb collisions at $\sqrt{S}=5.02$ TeV. The details of the calculation can be found in Ref.\,\cite{me-pa}. Our results is consistent with predictions given in Ref.\,\cite{me-pa}  at 4.4 TeV for minimum-bias collisions. Here, we quantify the theoretical errors  and provide predictions at 5.02 TeV at different centralities. The main input in the $k_T$ factorization for calculating the multiplicity is the dipole forward amplitude which we take the b-CGC saturation model \cite{b-cgc}. This model explicitly depends on the impact-parameter and approximately incorporates all known features of small-x physics and describes the small-x data at HERA including diffractive data \cite{b-cgc}, as well as RHIC and the LHC data at small-x \cite{me-pp,me-aa1,me-pa,me-aa}.  We have only two free parameters here, the mini-jet mass and the over-all normalization factor in \eq{PO1} which are fixed at lower energy\footnote{Note that the over-all normalization parameter absorbs three unknown factors,  the $K$-factor, a factor due to parton-hadron duality and a factor to relate the interaction area to the average impact-parameter of interaction.} \cite{me-pa}. Unfortunately, these two parameters cannot be uniquely fixed, because the error bars in experimental data for multiplicity at lower energy is rather large, and moreover,  there is a correlation between these two parameters. These uncertainties are carefully quantified  and shown in \fig{fig-m}. The same setup provides excellent description of  charged hadron multiplicity distribution in p+A collisions at RHIC at different centralities over a wide range of rapidity and also in p+p and A+A collisions at the LHC and lower energies \cite{me-pa,me-pp}.  

In \fig{fig-m}  (curves labeled by b-CGC), we show the charged hadron multiplicity in p+Pb collisions at 5.02 TeV for various centralities $0-20\%$, $20-40\%$, $40-60\%$, $60-80\%$  and minimum-bias, obtained from the $k_T$ factorization \eq{M1} using the b-CGC saturation model. The impact-parameter dependence of the saturation model is crucial here for defining the centrality of the collisions. In \fig{fig-m},  it is seen that the multiplicity distribution has the biggest asymmetry for more central collisions while for peripheral collisions like for example at $60-80\%$ centrality cut, the system becomes more similar to p+p collisions, and this fact is reflected in  the total charged hadron multiplicity distribution. This effect can be more clearly seen in left panel in \fig{fig-m}. Note that in all predictions shown in \fig{fig-m},  we assumed a fixed mini-jet mass $m_{jet}=5$ MeV for all energies/rapidity and centralities. As we already pointed out, the mini-jet mass is related to pre-hadronization/hadronization stage and cannot be obtained from saturation physics, and it was fixed via a fit to lower energy minimum-bias data. Unfortunately, for very peripheral collisions where the system becomes more like p+p collisions (or symmetric), this assumption is less reliable.  More importantly, one should also note that we used the $k_T$ factorization which is only proven for asymmetric collisions like p+A collisions at small-x, therefore for more peripheral collisions the current CGC prescription is less reliable. 

The upcoming experimental measurements of the charged hadron multiplicity at different centralities  at the LHC  can provide vital information about the impact-parameter dependence of the saturation scale. This can be also considered as a non-trivial test of underlying dynamics of gluon saturation, as the system at different centralities evolves from an asymmetric to more symmetric form and the total charged hadron multiplicity changes significantly, roughly speaking, up to a factor of about two compared to minimum-bias collisions, see \fig{fig-m}. 

The recent LHC data on the multiplicity in A+A collisions shows that the energy growth of multiplicities in A+A collisions is different from the corresponding p+p collisions \cite{e-lhc}. 
At first sight, the different power-law behavior of multiplicity in A+A and p+p collisions seems to be in conflict with the very idea of universality of particle production in the CGC/saturation picture. However, one should bear in mind that the $k_T$ factorization has been proven only at leading log approximation for a scatterings of a dilute partonic system on a dense one such as p+A collisions at high-energy \cite{kt,kt-rest}. On the contrary, it has been already shown that in the case of dense-dense collisions such as A+A central collisions it is not valid \cite{no-kt}. In Ref.\,\cite{me-aa1}, we showed that the gluon-decay cascade in pre-hadronization in the so-called 
MLLA (Modified Leading Logarithmic Approximation) regime which has a different kinematics compared to the BFKL type gluon emission, gives rise to extra energy-dependence. Although this should be still considered as initial-state effect, it is closely related to the open question of how 
to accommodate the fragmentation processes in the $k_T$ factorization formalism. In Ref.\,\cite{me-aa1}, we extracted gluon-jet decay cascade from $e^{+}e^{-}$ annihilation data and we showed that the energy-dependence of about $S^{0.036}$ due to gluon-decay cascade is exactly what explains the different power-law energy-dependence of hadron multiplicities in A+A compared to p+p collisions at the LHC. This effect is more important for A+A collisions at high energy where the saturation scale is larger and consequently the average transverse momentum of the jet becomes larger than 1 GeV. The MLLA gluon decay cascade enhances the hadron multiplicity about $10-25\%$ \cite{me-aa1}.

A comparison of various predictions coming from different saturation models for the charged hadron multiplicity in minimum-bias p+A collisions at 4.4 TeV, can be found in Ref.\,\cite{hard}. The upcoming p+A collisions at the LHC can in principle discriminate between different approaches and put more constrain on saturation models.

Next we provide our prediction for spectra of the single inclusive hadron and prompt photon production in terms of the nuclear modification factor\footnote{Upon request, we can also provide our theoretical predictions for the spectra of inclusive hadron production in p+A and p+p collisions at the LHC.} $R_{pA}$ defined as follows, 
\begin{eqnarray}
R^{ch}_{pA}&=&\frac{1}{N_{coll}}\frac{dN^{p A \rightarrow h X}}{d^2p_T d\eta}/
\frac{dN^{p p \rightarrow h X}}{d^2p_T d\eta},\nonumber\\
R^\gamma_{pA}&=&\frac{1}{N_{coll}}\frac{dN^{p A \rightarrow \gamma X}}{d^2p_T^\gamma d\eta_\gamma}/
\frac{dN^{p p \rightarrow \gamma X}}{d^2p_T^\gamma d\eta_\gamma},\
\end{eqnarray}
where $N_{coll}$ is the number of binary proton-nucleus collisions. We
take $N_{coll}=6.9$ in minimum-bias p+Pb collisions at $\sqrt{S}=5$  TeV \cite{ncoll}. In order to compare our $R_{pA}$ predictions with experimental data,
one needs to rescale our theoretical curves by matching the normalization $N_{coll}$
to the experimental value. We will use the NLO MSTW 2008 PDFs \cite{mstw} and the NLO KKP FFs \cite{kkp} for neutral pion and charged hadrons. For the photon fragmentation function, we will use the full leading log parametrization \cite{own,ffp}.  We assume the factorization scale $Q$ in the FFs and the PDFs to be equal. In order to investigate the uncertainties associated with the choice of the hard-scale $Q$, we consider various cases of $Q=2p_T,p_T, p_T/2$  and $Q=2p^\gamma_T,p^\gamma_T, p^\gamma_T/2$ for inclusive hadron production  \eq{final} and  prompt photon production Eqs.\,(\ref{pho2},\ref{pho4}), respectively.

Solving the rcBK equation (\ref{bk1}) in the presence of the impact-parameter is still an open problem. However, for the minimum-bias analysis considered here the impact-parameter dependence may be less important. Then, the initial saturation scale $Q_{0s}$ should be considered as an impact-parameter averaged value and it is extracted from the minimum-bias data. The main input in the rcBK evolution equation is the initial dipole profile \eq{mv} with two unknown parameters $Q_{0s}$ and $\gamma$.  The current HERA and RHIC data alone are not enough to uniquely fix the values of $Q_{0p}$ and $\gamma$, namely there exists several parameter sets of ($Q_{0p}$, $\gamma$) which all provide a good description 
of the data\footnote{This is partly because there is indeed correlation between parameters of the model, and in order to constrain the model, more exclusive data at larger kinematic regions are required. In this sense, the upcoming LHC data on the p+A collisions will be very useful.} \cite{jav1}. However, the recent LHC data on p+p collisions for spectra of the inclusive hadron production  provided more constrains on the  values of  $Q_{0p}$ and $\gamma$. It was recently shown that  $Q_{0p}^2 \approx 0.168\,\text{GeV}^2$ with $\gamma \approx 1.119$ provides a good description of the small-x data at the LHC, HERA and RHIC with a proton target \cite{jav1,j1,j2}, see also Ref.\,\cite{ja1}.  We will take these values for the rcBK description of the projectile proton.

Having known the rcBK description of the proton, now we should determine the rcBK description of a nucleus. In the CGC picture, at high-energy or small-x, there is no difference between a proton and nucleus except in their saturation scale. Obviously in the case of the nucleus, the impact-parameter dependence is more important. Given that the perturbative rcBK evolution equation does not incorporate non-perturbative confinement effect \cite{bk-c} and the b-dependent solution of the BK equation is not yet known, it seems like we are doomed to go beyond the current CGC framework and add several model assumptions. Here, we stay in the standard CGC framework without invoking any extra ingredient to the model. Our only assumption is that like in the case of proton, the initial saturation scale of a nucleus  $Q_{0A}$ is an impact-parameter averaged value and can  be extracted from other reactions with nuclear targets. In our approach, the role of the fluctuations on particle production and all other possible non-perturbative effects are effectively incorporated into the average value of $Q_{0A}$ extracted from experimental data. Then we assume that the small-x evolution (the BK or JIMWLK equations) will not be altered due to non-perturbative soft physics. For the minimum-bias collisions, it is generally assumed that the initial saturation scale of a nucleus with atomic mass number A, scales linearly with $A^{1/3}$ \cite{mv,cgc-review1,al}, namely we have 
\begin{equation}\label{sa0}
Q_{0A}^2=c A^{1/3}~Q_{0p}^2, 
\end{equation}
where the parameter $c$ can be fixed via a fit to data. In Ref.\,\cite{raju}, it was shown that  New Muon Collaboration's (NMC) data at small-x can be described with $c\approx 0.5$.  This is also consistent with the fact that  RHIC inclusive hadron production data in minimum-bias deuteron-gold (d+A) collisions and DIS data for heavy nuclear targets can be described with  $Q_{0A}^2\approx 3 \div 4~Q_{0p}^2$  \cite{ja1,urs,raju,jav-r}. However, one should bear in mind that  RHIC and DIS data on heavy nuclear target are very limited in number of data points with rather large errors. Most importantly, RHIC data in d+A  collisions at forward rapidities and nuclear DIS data at small-x are limited in kinematics to low transverse momenta (and low virtualities), about $p_T<4$  GeV. At high $p_T$, we expect that $R^{ch}_{pA}\to 1$. Assuming that in high-energy collisions at high-$p_T$ ($p_T>4 $ GeV) we are still in the saturation region (which can be the case at the LHC), then by making use of the $k_T$ factorization or hybrid formalism, at high-$p_T$ we approximately have, 
\begin{equation}
R^{ch}_{pA} (p_T >>1)= \frac{Q_{0A}^2 S_A}{Q_{0p}^2 A S_p}\approx \frac{Q_{0A}^2}{Q_{0p}^2 A^{1/3}} \to 1, 
\end{equation}
where $S_{A}$ and $S_p$ are effective interaction area in p+A and p+p collisions, respectively. This leads to the same relation as in \eq{sa0}, but with $c\approx 1$. Therefore,  for heavy nuclei (with $A\approx 208$), the initial nuclear saturation scale can vary within: 
\begin{equation} \label{qa}
 3 \div 4~Q_{0p}^2 \le Q_{0A}^2 (x_0=0.01) \le 6\div 7\,Q_{0p}^2.
\end{equation}
We recall that the initial saturation scale of proton extracted from HERA data is rather small $Q_{0p}^2=0.168\,\text{GeV}^2$. Therefore, the upper limit of  the initial saturation scale of lead nucleus is about $Q_{0A}^2 (x_0=0.01) \le 1.2\,\text{GeV}^2$ which is consistent with phenomenological saturation models based on RHIC and HERA data, 
see Ref.\,\cite{cgc-review1} and references therein. In \eq{qa}, it seems as if we made an assumption at which value of $x$ we will have $R^{ch}_{pA}\to 1$. However, the assumed value of $x_0$ will not affect the inequality given in \eq{qa}  in saturation region. In other words, because the speed of the rcBK evolution in rapidity is the same for both proton and nucleus, the inequality given in \eq{qa} should be also valid at $x_0=0.01$.  In previous studies of the nuclear modification at the LHC \cite{ja1,jav-r,raju-rpa}, in obtaining the rcBK equation solutions, only lower limit of the initial saturation scale for nucleus was considered. Here we will extend our analysis by considering entire range of $Q_{0A}$ given in \eq{qa}.

\begin{figure}[t]       
                               \includegraphics[width=6.9 cm] {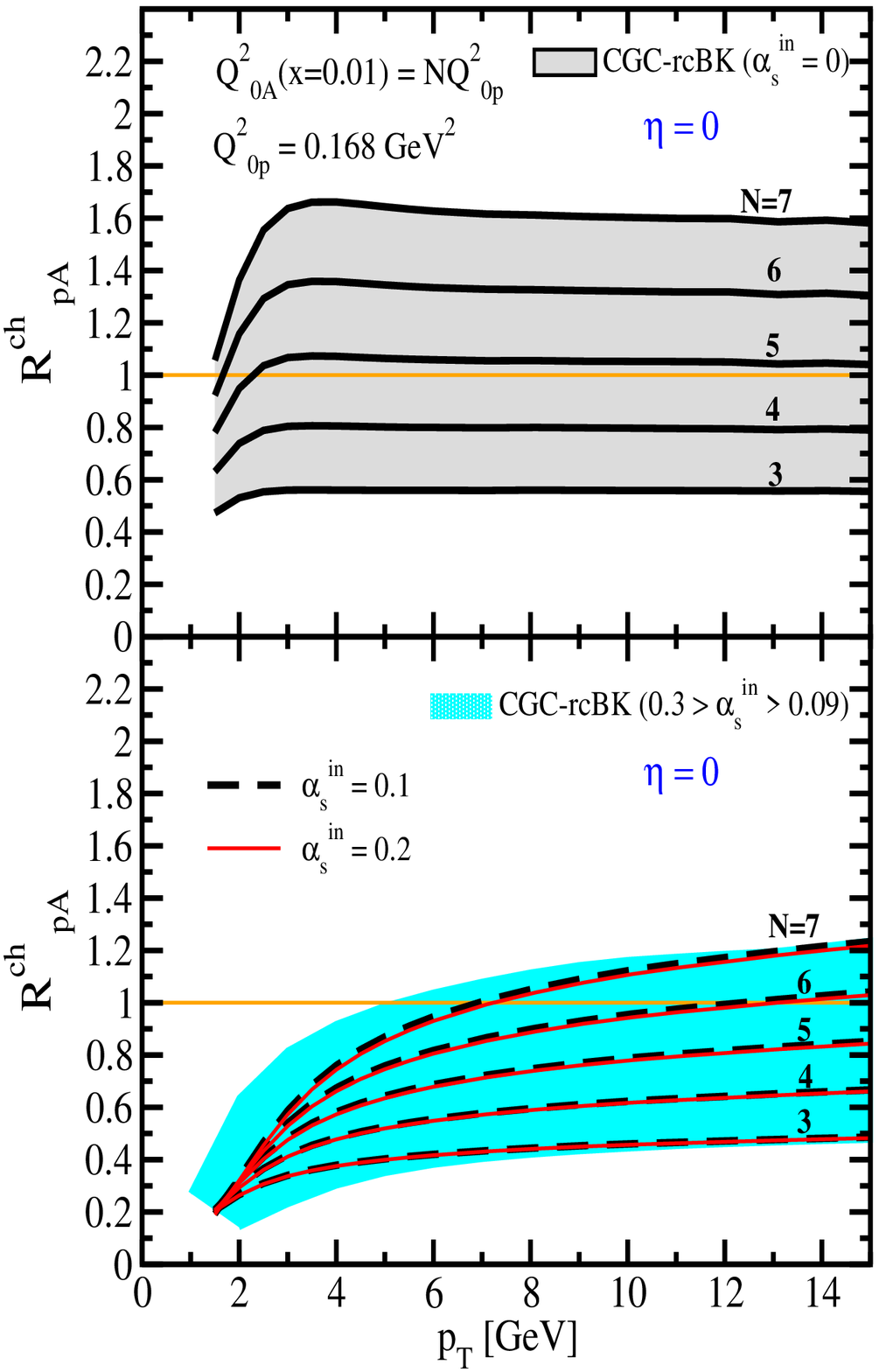}       
                                \includegraphics[width=6.9 cm] {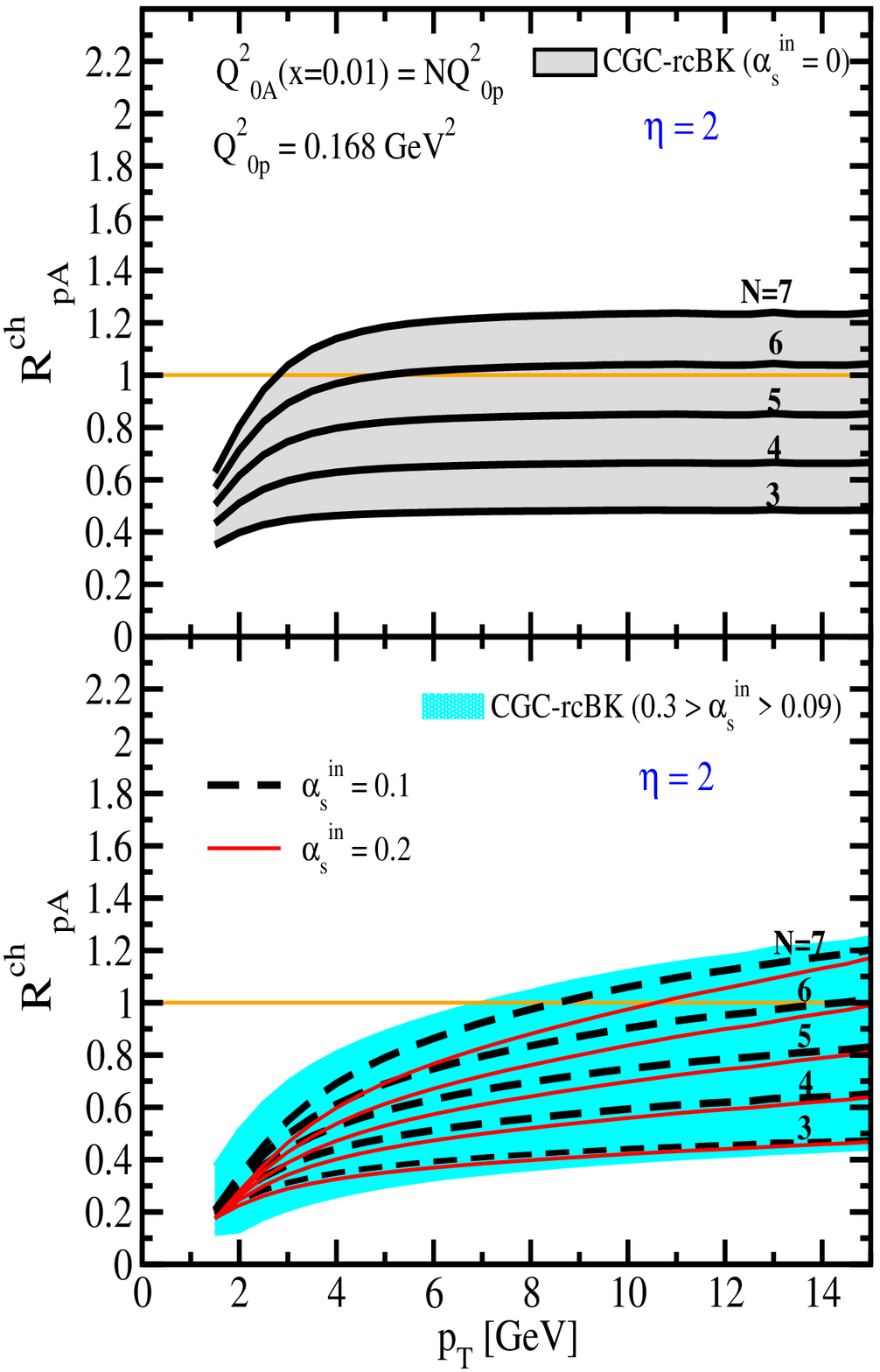}      
                                \includegraphics[width=6.9 cm] {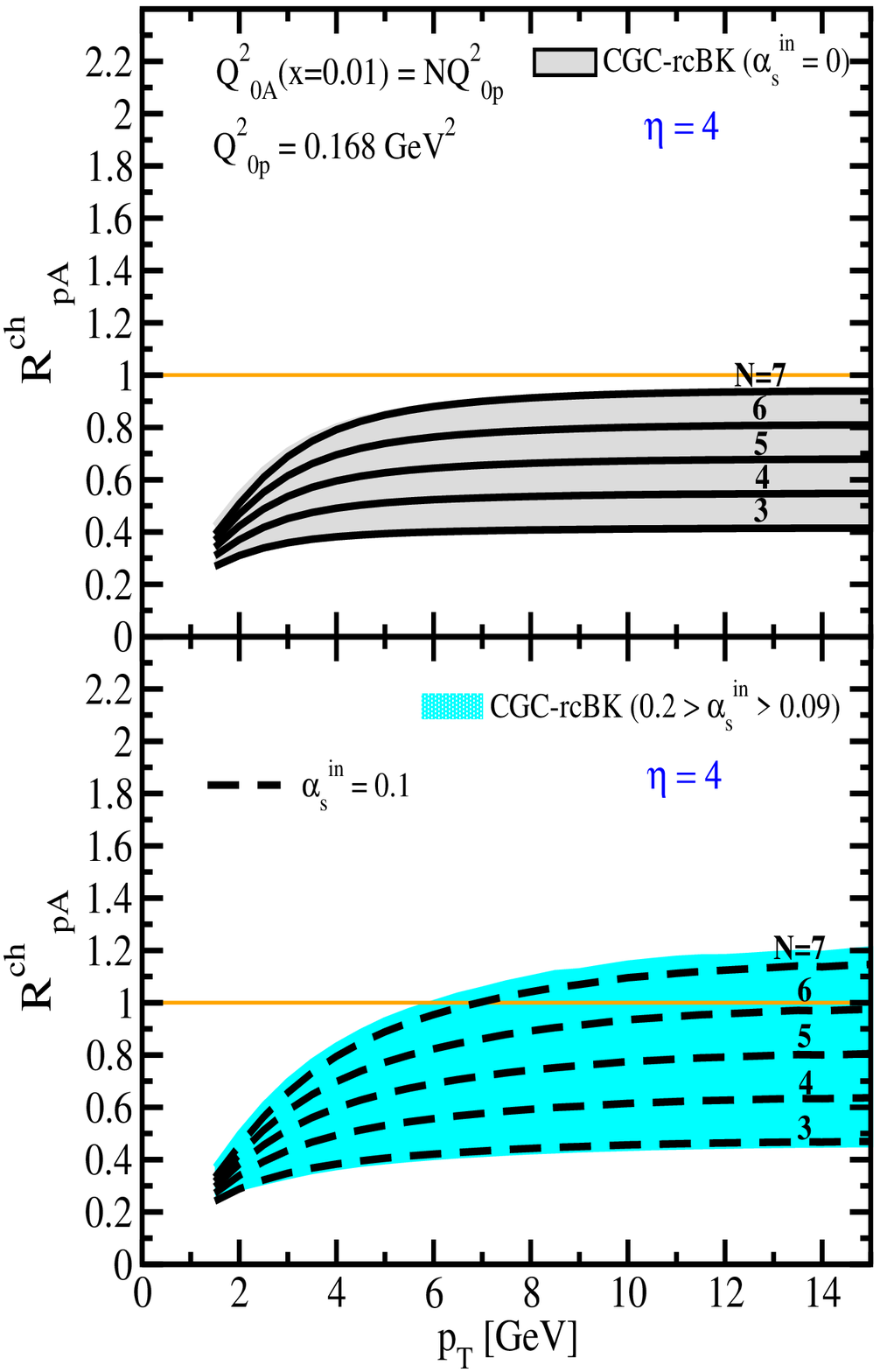}       
                                \includegraphics[width=6.9 cm] {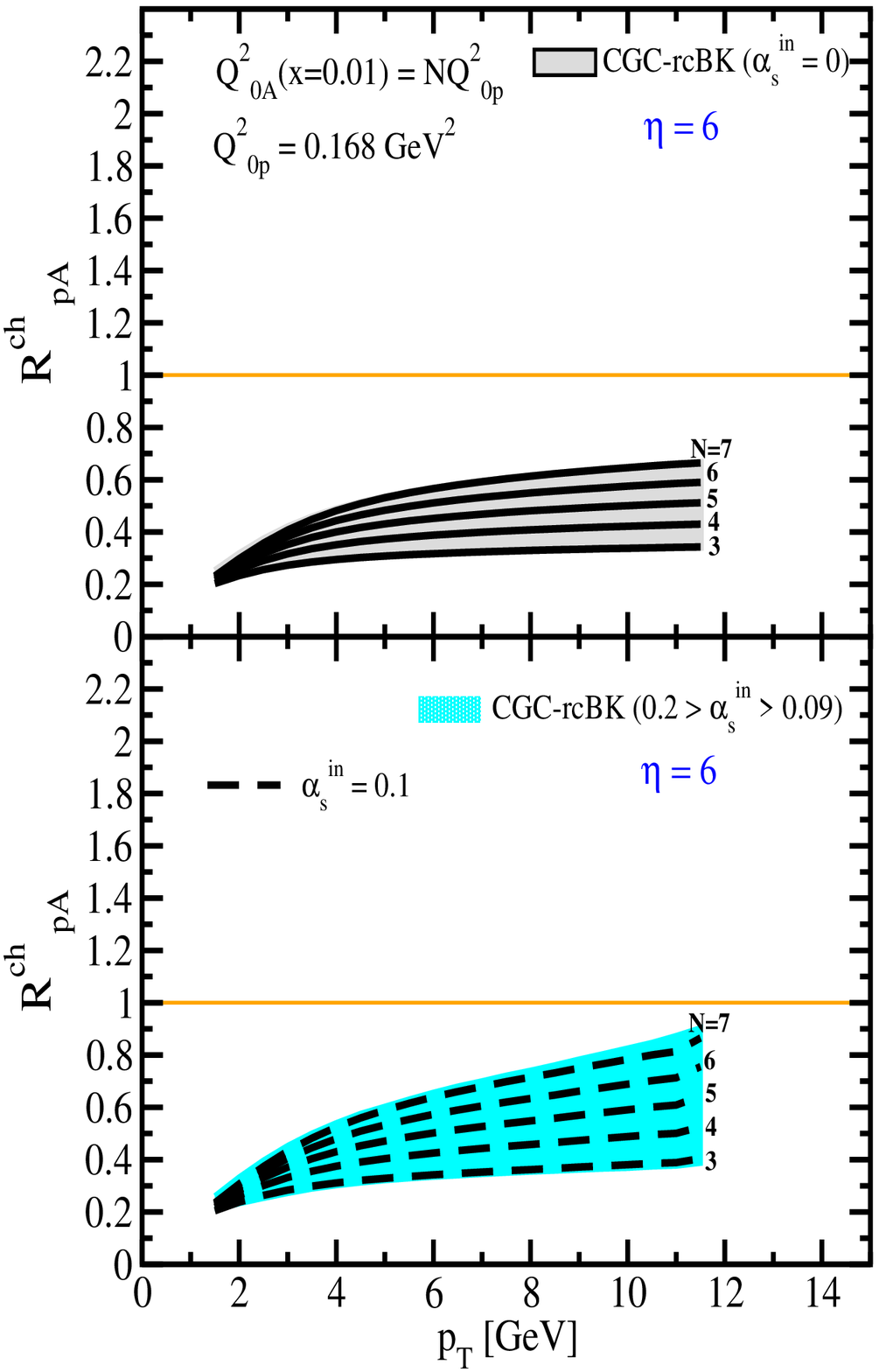}     
\caption{ The nuclear modification factor $R_{pA}^{ch}$ for inclusive charged hadrons $h^{+}+h^{-}$  production in minimum-bias p+Pb collisions at $\sqrt{S}=5$ TeV at different rapidities $\eta=0,2,4,6$
               obtained from the hybrid factorization \eq{qa}  with the solutions of the rcBK with different initial saturation scale for nucleus.  The band labeled CGC-rcBK includes uncertainties due to the variation of the initial saturation scale of nucleus and different factorization scale $Q$. At every rapidity, we also show the  results  by taking  $\alpha_s^{in}=0$ (only elastic contribution) and $0.3\div 0.2\le \alpha_s^{in}\le 0.1$. The lines labeled by a number $N$ are the results with a fixed  hard-scale $Q=p_T$ and a fixed saturation scale $Q_{0A}^2=N Q_{0p}^2$  with $N=3\div 7$ constrained in \eq{qa} and $Q_{0p}^2=0.168\,\text{GeV}^2$. }
\label{rp-h}           

\end{figure}

\begin{figure}[t]       
                               \includegraphics[width=7. cm] {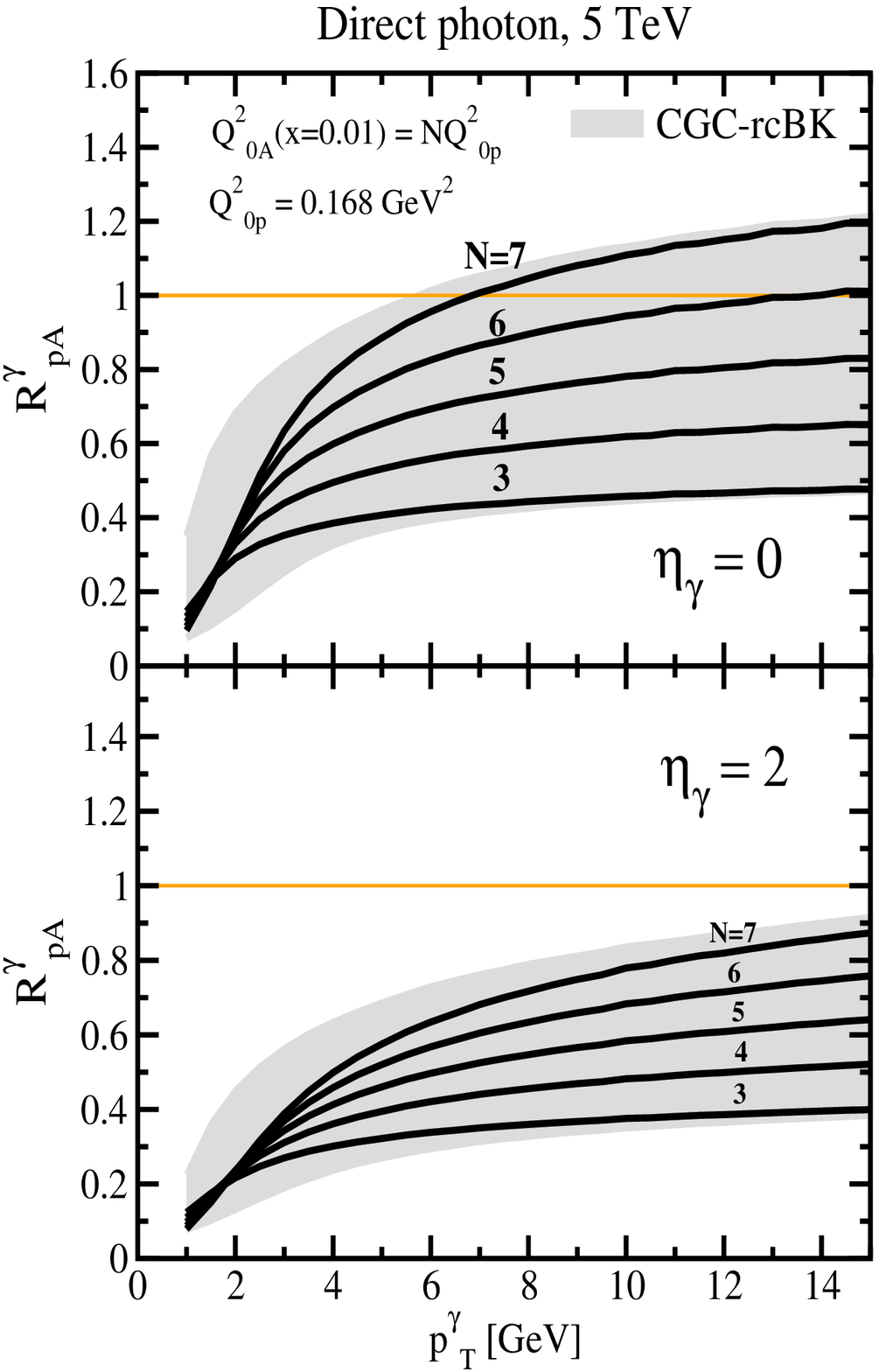}       
                                \includegraphics[width=7. cm] {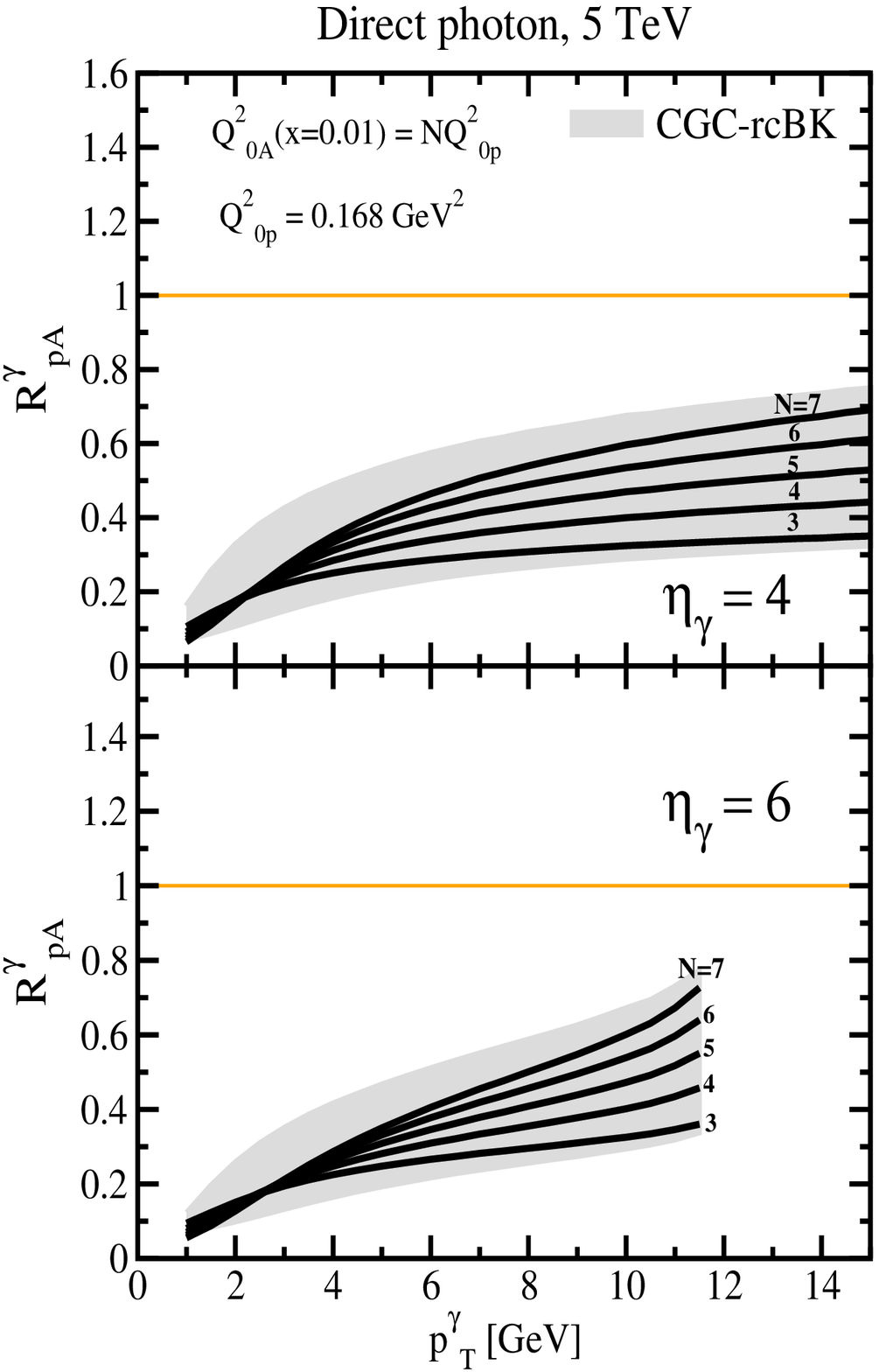}      
     
\caption{The nuclear modification factor $R_{pA}^{\gamma}$ for direct photon production in minimum-bias p+Pb collisions at $\sqrt{S}=5$ TeV at different rapidities $\eta_{\gamma}=0,2,4,6$
               obtained from \eq{pho4}  with the solutions of the rcBK with different initial saturation scale for nucleus.  The band labeled CGC-rcBK includes uncertainties due to the variation of the initial saturation scale of nucleus and different factorization scale $Q$.  Similar to \fig{rp-h}, the lines labeled with a number $N$ are the results with a fixed  hard-scale $Q=p^{\gamma}_T$ and a fixed saturation scale $Q_{0A}^2=N Q_{0p}^2$  with $N=3\div 7$ constrained in \eq{qa} and $Q_{0p}^2=0.168\,\text{GeV}^2$.     }
\label{rp-p}
\end{figure}
\begin{figure}[t]       
                                   \includegraphics[width=7. cm] {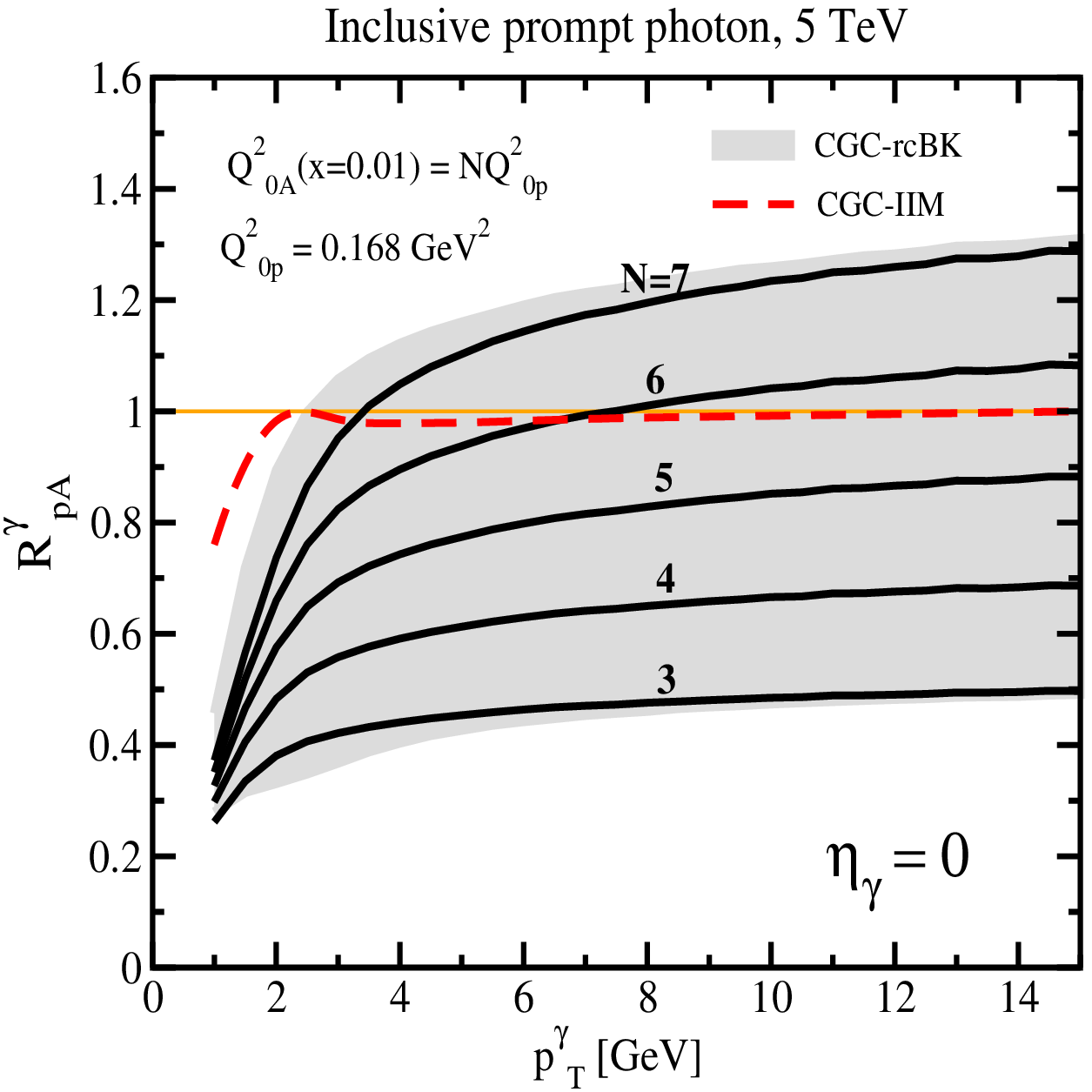}      
                                \includegraphics[width=7. cm] {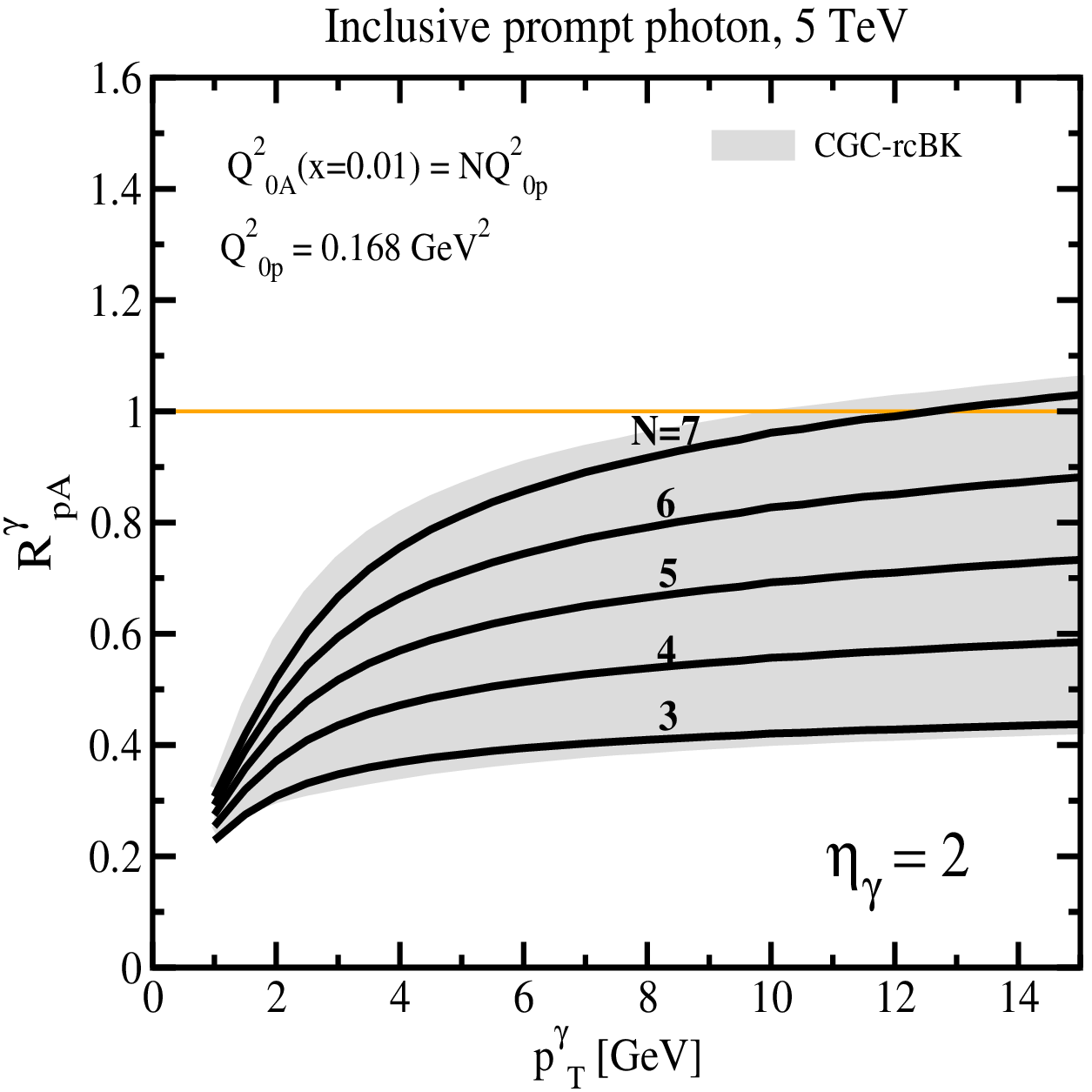}      
     
\caption{The nuclear modification factor $R_{pA}^{\gamma}$ for single inclusive prompt photon production in minimum-bias p+Pb collisions at $\sqrt{S}=5$ TeV at different rapidities $\eta_{\gamma}=0,2$. The descriptions of  the band and the solid black lines are the same as in \fig{rp-p}.  The dashed red line (labeled CGC-IIM) is calculated from the same master \eq{cs} but a different saturation model, namely the so-called  Iancu-Itakura-Munier (IIM) saturation model \cite{iim,me-nosh}.   }
\label{rp-p2}
\end{figure}

In \fig{rp-h}, we show the nuclear modification  $R_{pA}^{ch}$ for inclusive charged hadron production in minimum-bias p+Pb collisions at $\sqrt{S}=5$ TeV at different rapidities $\eta=0,2,4,6 $ obtained from the hybrid factorization \eq{final} supplemented with the rcBK evolution equation. The band labeled CGC-rcBK includes uncertainties associated with the variation of initial nuclear saturation scale  within the range given in \eq{qa} in obtaining the rcBK equation solutions, and also the uncertainties coming from various choices of the factorization scale $Q=2p_T, p_T, p_T/2$  in \eq{final}. It was already shown in \cite{ja1} that unfortunately with RHIC data alone, one cannot fix the value of  $\alpha_s^{in}$ in \eq{final}, and with a reasonable $K$-factor any value of $\alpha_s^{in}$ within $0.3 \div 0.2 \le \alpha_s^{in}\le 0$ can be consistent with RHIC inclusive hadron production data both in p+p and p+A collisions. In \fig{rp-h}, we show the effect of various values for the strong-coupling $\alpha_s^{in}$ in \eq{final}, namely in every panel for a given rapidity we also show the results with $\alpha_s^{in}=0$ and  $0.3\div 0.2\le \alpha_s^{in}\le 0.1$.  In \fig{rp-h}, it is generally seen that inelastic contribution to the hybrid formalism distorts the  $R_{pA}^{ch}$ flatness with transverse momentum, namely at low-$p_T$ for more central collisions it leads to more suppressions while for more forward collisions it enhances $R_{pA}^{ch}$ at high $p_T$. This behavior was numerically first observed in Ref.\,\cite{ja1}.  It is interesting to notice that for more central collisions, let's say at $\eta=0, 2$, where the higher-order corrections are important, the inelastic contribution becomes more sensitive to the saturation effect and it leads to more suppression at low $p_T$. This is more obvious for a  bigger initial nuclear saturation scale at low-$p_T$. This is in accordance with the fact that inelastic terms are generally more sensitive to the saturation physics. This expectation was first qualitatively pointed out in Ref.\,\cite{inel}.

In order to better quantify the effect of uncertainties due to different initial nuclear saturation within the constrained range given in \eq{qa}, in \fig{rp-h}, we show $R_{pA}^{ch}$ obtained from rcBK solutions with different initial nuclear saturation scale $Q_{0A}^2=N Q_{0p}^2$  with $3\le N \le 7$, and with factorization scale taken $Q=p_T$. 
In order to demonstrate uncertainties associated with various value for the strong-coupling $\alpha_s^{in}$  in \eq{final}, we also show $R_{pA}^{ch}$ for fixed values  $\alpha_s^{in}=0, 0.1, 0.2$. Unfortunately, $R^{ch}_{pA}$ is very sensitive to the initial saturation scale of nucleus, and variation of $Q_{0A}$ in range given in  \eq{qa} gives rise to rather large uncertainties for $R^{ch}_{pA}$. However,  if the experimental measurement of $R^{ch}_{pA}$ at one rapidity is performed then one can use this data to extract the initial nuclear saturation scale from Figs.\,(\ref{rp-h},\ref{rp-p}).  Let's assume that experimental value of  $R_{pA}^{ch}$ at one rapidity, let's say at $\eta=2$,  lies between two lines labeled by  $N_1, N_2$, then for other rapidities, our predictions between two lines with the same labeling $N_1, N_2$ which has the same $Q_{0A}$, should be only considered as true predictions of the CGC. In this way, by knowing $R_{pA}^{ch}$ at one rapidity, one can significantly  reduces theoretical uncertainties associated with $Q_{0A}$ at other rapidities. Note that our predictions here are consistent with earlier studies in \cite{ja1,j2,raju-rpa}. It is generally seen that the CGC predicts more suppression at forward rapidities compared to the collinear factorization results \cite{co1}. Moreover, the small-x evolution washes away the Cronin-type peak  at low-$p_T$ at all rapidities. Therefore, observation of a strong Cronin-type peak in p+Pb  collisions at the LHC, regardless if it is enhancement or suppression, can be potentially a signal of non-CGC physics, and it will be difficult to accommodate this feature in the CGC approach at current accuracy\footnote{This was also pointed out in the talks given by Xin-Nian Wang and Adrian Dumitru in pA@LHC workshop,  CERN, Geneva,  June 2012.}, see also Refs.\,\cite{me-nosh,non1,non2,non3}.

In Figs.\,(\ref{rp-p},\ref{rp-p2}), we show our predictions for the nuclear modification $R_{pA}^{\gamma}$ of direct photon and single inclusive prompt photon production in minimum-bias p+Pb collisions at 5 TeV at various rapidities obtained from Eqs.\,(\ref{pho2},\ref{pho4}) with solutions of the rcBK with different initial saturation scale for nucleus. Similar to the case of inclusive hadron production, $R_{pA}^{\gamma}$ for direct photon production (and inclusive prompt photon production) is also very sensitive to the initial saturation scale of nucleus. The band labeled by CGC-rcBK includes uncertainties due to the variation of the initial saturation scale of nucleus and different factorization scale $Q=2p_T^\gamma, p_T^\gamma, p_T^\gamma/2$.  We also show our predictions for $R_{pA}^{\gamma}$ with various $Q_{0A}^2=N Q_{0p}^2$ and $N=3\div 7$ constrained in \eq{qa}.  Again, the value of $N$ can be extracted from $R_{pA}^{\gamma}$ or $R_{pA}^{ch}$ measurements at one rapidity using the predictions shown in Figs.~(\ref{rp-h},\ref{rp-p},\ref{rp-p2}) for different values of $N$.  Then at other rapidities the corresponding curves labeled with $N$ should be taken as our prediction. It is seen from Figs.\,(\ref{rp-p},\ref{rp-p2}) that $R_{pA}^{\gamma}$ of direct photon and inclusive prompt photon are rather similar, although
the direct photon production is generally more suppressed than the inclusive prompt photon production. This is what one may expect in our picture since direct photon cross-section in \eq{pho2} probes the target structure function at lower transverse momentum $p_T^\gamma$ and consequently lower x than the fragmentation part. 

In \fig{rp-p2}, at $\eta_\gamma=0$ we also show  $R_{pA}^{\gamma}$ for the inclusive prompt photon production calculated in \cite{me-nosh}  by Iancu-Itakura-Munier (IIM) saturation model \cite{iim}. In the IIM model, the saturation is approached from the BFKL region, and therefore the small-x evolution encoded in this model is different from the rcBK equation. Note that the IIM saturation model also provide a good description of HERA data \cite{iim}. Nevertheless, it is seen from \fig{rp-p2} that depending on the value of $Q_{0A}$, the rcBK and the IIM saturation model provide different suppression for inclusive prompt photon production in minimum-bias p+A collisions at the LHC\footnote{Note that in Ref.\,\cite{me-nosh}, the coordinate representation of \eq{cs} was used. But as it was shown in Ref.\,\cite{pho-cgc2}, this is equivalent to the CGC formulation given in \eq{cs}, see also Refs.\,\cite{boris,me-pho-old}.}.  Note that similar to the inclusive hadron production, the CGC approach generally gives larger suppression for inclusive prompt photon $R_{pA}^{\gamma}$ at forward rapidities compared to collinear factorization results \cite{co2}. 

\begin{figure}[t]       
                              \includegraphics[width=7 cm] {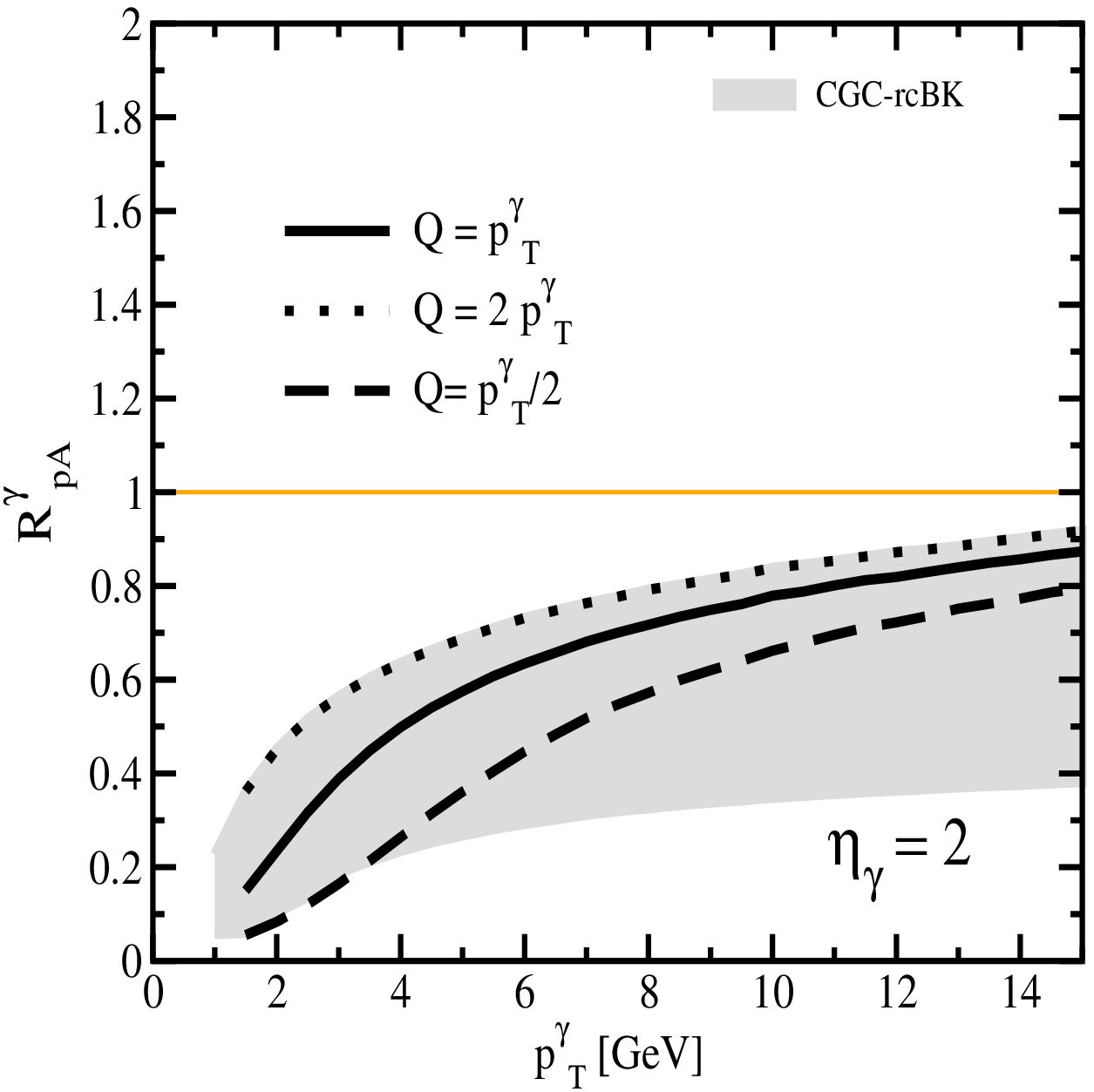}        
                            \includegraphics[width=7 cm] {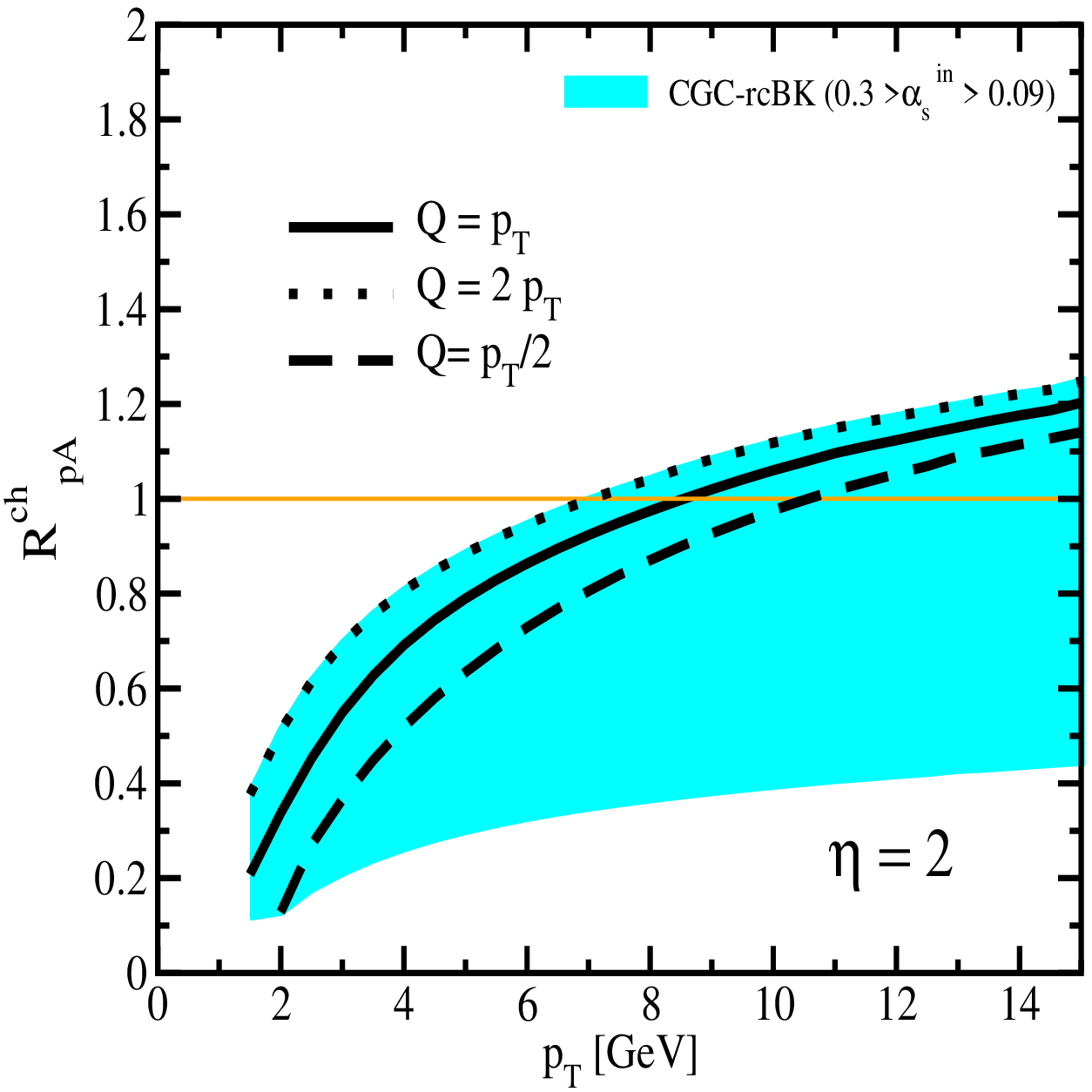}     
\caption{ The nuclear modification factor in minimum-bias p+Pb collisions at the LHC $\sqrt{S}=5$ TeV  and $\eta=\eta_\gamma=2$ for inclusive hadron production (right panel) and direct photon production (left panel) with different hard-scale (or factorization scale) $Q$ in Eqs.~(\ref{pho4},\ref{final}). The initial saturation scale for nucleus is fixed for all lines with $N=7$. The bands are the same as in Figs.\,(\ref{rp-h},\ref{rp-p}).}
\label{rp-f}
\end{figure}

In \fig{rp-f}, we show the effect of different choices for the hard-scale $Q$ appeared in the factorization formulas Eqs.~(\ref{final},\ref{pho4}) for both inclusive hadron production (right panel) and direct photon production (left panel) at forward rapidity $\eta=\eta_\gamma=2$. The initial nuclear saturation scale is fixed for all lines. Note that in the case of inclusive hadron production with $\alpha_s^{in}=0$ (only elastic terms), the hard-scale $Q$ only appears in the PDFs and FFs, and the effect of different value for $Q$ is negligible in  the $R_{pA}^{ch}$. However, in the presence of inelastic contribution ($\alpha_s^{in}\ne 0$), the choice of hard-scale $Q$ becomes important and it leads to a sizable effect for the $R_{pA}^{ch}$, see \fig{rp-f}.  The sensitivity  of the nuclear modification factor $R^{\gamma}_{pA}$ and $R^{ch}_{pA}$ to the hard-scale $Q$ (and the strong-coupling $\alpha_s^{in}$) clearly indicates that the higher-order corrections should be important. Note that for the case of inclusive hadron production, the full NLO corrections to the hybrid formalism has been recently calculated \cite{nlo-kt}, but yet to be employed for phenomenological purpose. 

Some words of caution are in order here. Our formulation is valid for asymmetric collisions when a projectile can be treated in the standard collinear approximation while for the target we systematically incorporated the small-x re-summation (at the leading twist approximation) effects.  Note, however, our reference for $R_{pA}^{ch}$ or $R_{pA}^{\gamma}$ is p+p collisions where at mid-rapidity and low transverse momentum the interacting system becomes symmetric, and therefore the CGC formulation (both the $k_T$ factorization and the hybrid formalism) will be less reliable. Moreover, our parameter sets for the rcBK equation was obtained from a fit to HERA data at small-x  $x<0.01$ and for low virtualities $Q^2\in[0.25,40] \,\text{GeV}^2$ \cite{jav1}. Therefore, our predictions at high-$p_T$ ($p_T,p_T^{\gamma}> 6\div 7$ GeV) should be taken with a grain of salt. 

\begin{figure}[t]       
                              \includegraphics[width=7 cm] {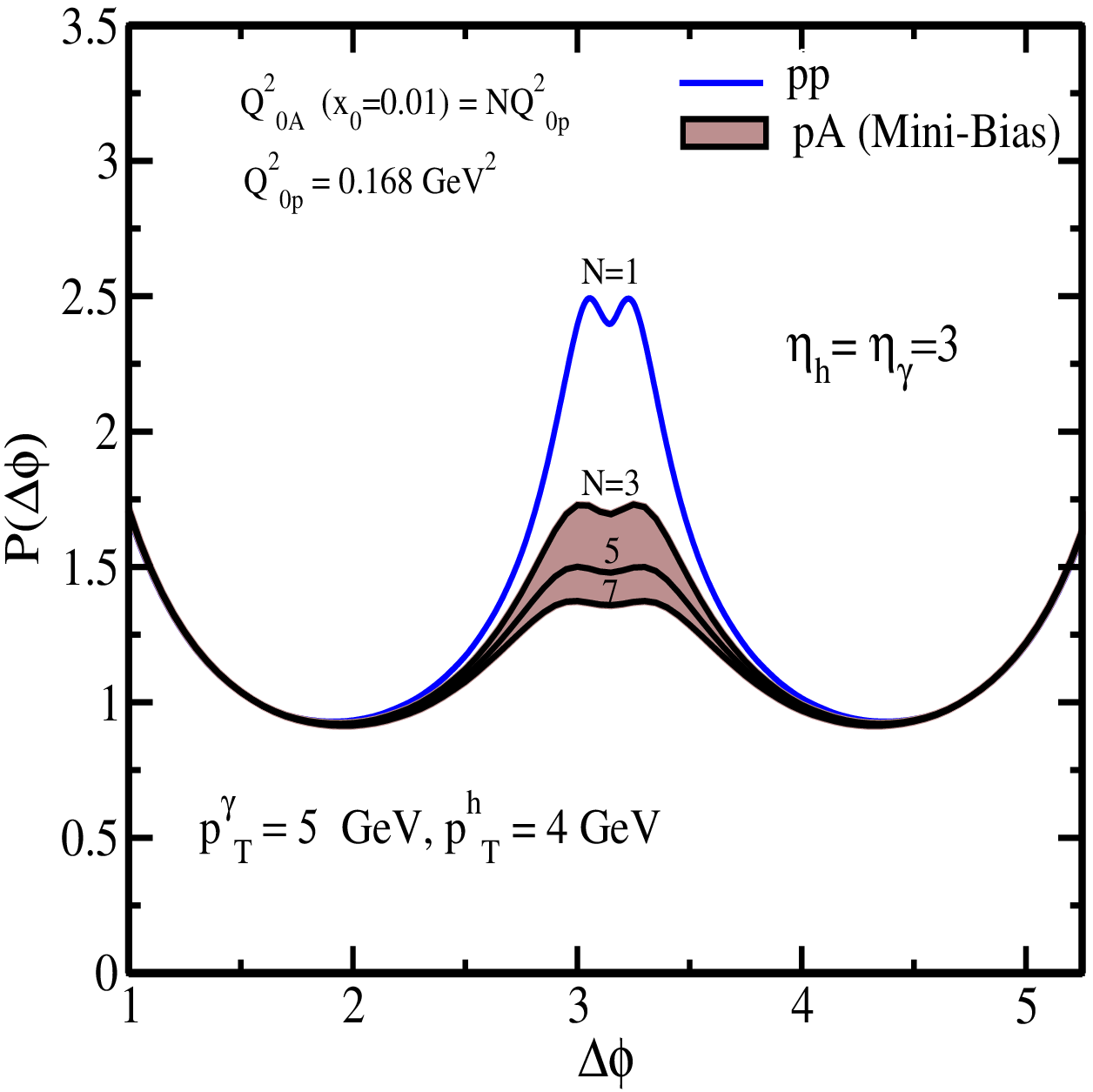}        
                            \includegraphics[width=7 cm] {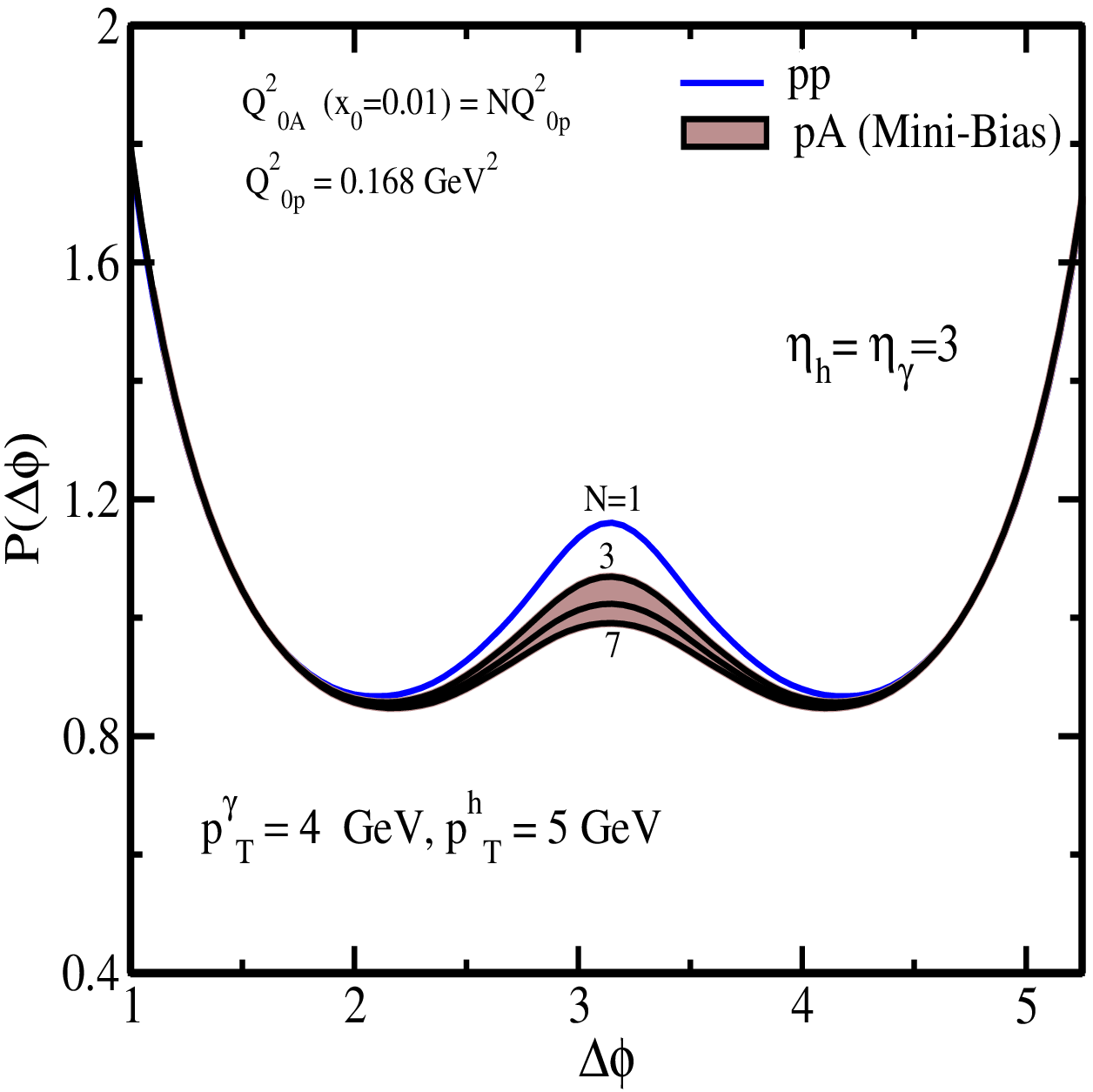}     
\caption{The photon-hadron correlation $P(\Delta \phi)$ defined in \eq{az} in minimum-bias (Mini-Bias) p+p and p+Pb collisions at $\sqrt{S}=5$ TeV and $\eta_{h}=\eta_{\gamma}=3$ obtained via the rcBK evolution equation with different initial nuclear saturation scale $Q^2_{0A}=NQ^2_{0p}$ and $N=3,5,7$ for two different transverse momentum regions $p^h_T>p^{\gamma}_T$ (right) and $p^h_T< p^{\gamma}_T$ (left). }
\label{rp-c}
\end{figure}

Finally, in \fig{rp-c}, we show our predictions for the azimuthal photon-hadron correlations in minimum-bias p+p and p+Pb collisions at 5 TeV for two different kinematics, namely in 
right panel $p^h_T>p^{\gamma}_T$ and in left panel $p^h_T< p^{\gamma}_T$. The azimuthal photon-hadron correlation is defined as \cite{ja2,me-ph}, 
\begin{equation}\label{az}
P(\Delta \phi)={d\sigma^{p\, A \rightarrow h(p_T^h)\,\gamma(p_T^\gamma)\, X}
\over d^2\vec{b_t}\,p_T^h dp_T^h\, p_T^\gamma dp_T^\gamma\, d\eta_{\gamma}\, d\eta_h\, d\phi} [\Delta \phi]/ {d\sigma^{p\, A \rightarrow h(p_T^h)\,\gamma(p_T^\gamma)\, X}
\over d^2\vec{b_t}\,p_T^h dp_T^h\, p_T^\gamma dp_T^\gamma\, d\eta_{\gamma}\, d\eta_h\, d\phi} [\Delta \phi= \pi/2],
\end{equation}
where the photon-hadron cross-section is given in \eq{qh-f}.  The function $P(\Delta \phi)$ gives the probability of the semi-inclusive photon-hadron pair production at a certain kinematics and angle  $\Delta \phi$, triggering the same production with the same kinematics at a fixed reference angle $\Delta \phi_c=\pi/2$. 
 It is seen that given the transverse momenta of the produced photon and hadron, the photon-hadron correlation can have a double or single peak structure. In Ref.\,\cite{me-ph} it was shown that this feature is related to saturation physics and it is mainly controlled by the ratio $p_T^h/p_T^\gamma$ (as can be seen in \fig{rp-c}). We refere the interested readers to Ref.\,\cite{me-ph} for the details and discussions. Here we only stress that the photon-hadron azimuthal correlations in p+A and p+p collisions can be considered as an excellent probe of the small-x dynamics \cite{ja2,me-ph}.  In \fig{rp-c}, we show the sensitivity of the correlations to various initial saturation scale.  Although the uncertainties associated to the initial saturation scale changes the strength of the correlation, 
remarkably the main features of the photon-hadron correlations which are intimately connected to saturation physics, seem to be robust, namely the suppression of the away-side correlations in p+A compared to p+p collisions, and appearance of the double or single peak structure. The predictions for the azimuthal angle correlations between the produced prompt photon and hadron calculated via the coincidence probability in p+A collisions at $\sqrt{S}=5$ TeV  can be found in Ref.\,\cite{me-ph}.

To summarize, in this letter within the CGC/saturation framework, we provided various predictions for different observables for the upcoming p+Pb collisions at $\sqrt{S}=5$ TeV at various rapidities.  The main caveats of the current CGC approach are twofold, lack of systematic control of higher order corrections and lack of enough high-quality experimental data at small-x to constrain the initial dipole profile of the rcBK evolution equation. 
We showed that the nuclear modification factor for inclusive hadron and direct photon production at the LHC are sensitive to both caveats. 
 Unfortunately, the current small-x data on heavy nuclei cannot uniquely fix the dipole parameters and the initial nuclear saturation scale $Q_{0A}(x=0.01)$. This gives rise to rather sizable theoretical uncertainties for $R_{pA}$. We made detailed predictions for different observables in p+Pb collisions at the LHC  with various values of $Q_{0A}$ (labeled by $N$) constrained by existing experimental data at small-x. If experimental data at a given rapidity for observables considered here are know, one can extract $Q_{0A}$ (or $N$)  by confronting our predictions in Figs.~(\ref{rp-h},\ref{rp-p},\ref{rp-p2},\ref{rp-c})  with experimental data, then at other rapidities our results labeled with the corresponding $N$ should be only considered as true CGC predictions. In this way, despite our rather large theoretical uncertainties, one can still test the main dynamics of the CGC/saturation at the LHC.

\begin{acknowledgments}
The author would like to thank Adrian Dumitru,  Jamal Jalilian-Marian, Genya Levin, Raju Venugopalan and Bo-Wen Xiao for useful discussions. 
This work is supported in part by Fondecyt grant 1110781. The author acknowledges the use of the USM HPC cluster provided by project BASAL FB0821.
\end{acknowledgments}


\section*{Note added 29 Nov. 2012: \small{Comparing our predictions with ALICE data} }

\begin{figure}[t]                                                            
                             \includegraphics[width=7 cm] {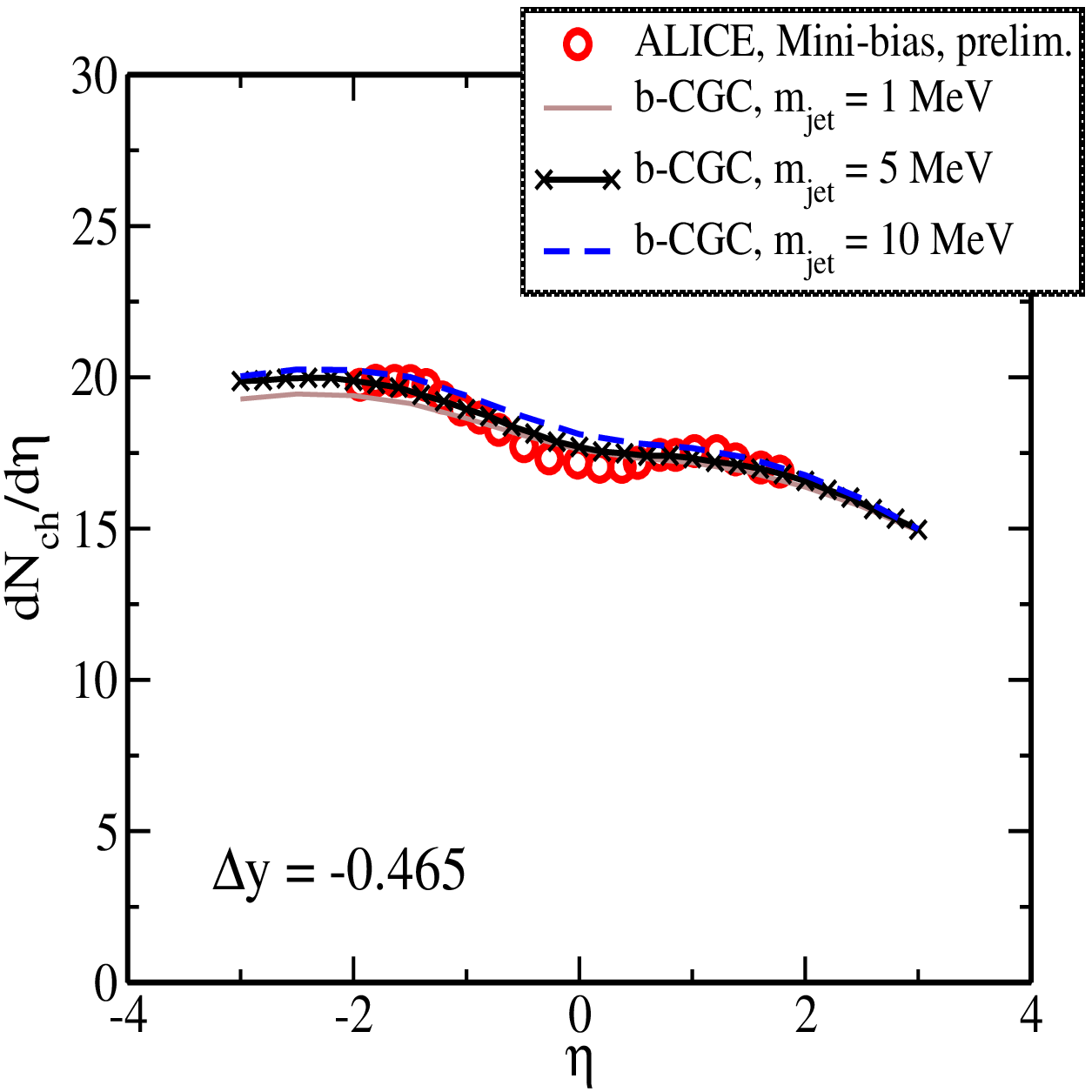}                 
                              \includegraphics[width=7 cm] {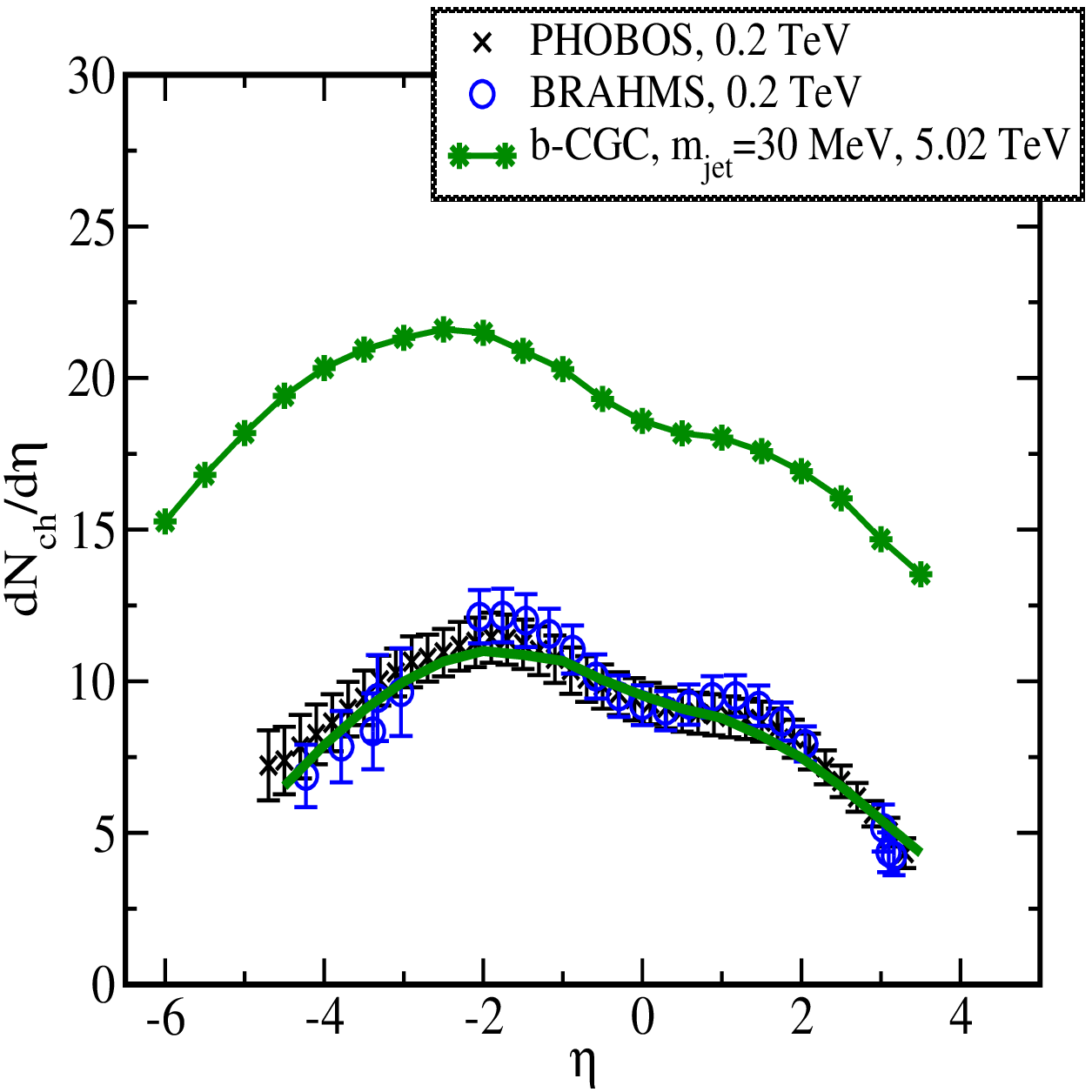}                                                    
\caption{Pseudorapidity distribution of the
                                  charged particles production in minimum-bias (Mini-bias) p+A
                                   collisions at the LHC
                                  $\sqrt{S}=5.02$ TeV. The theoretical
                                  curves labeled by b-CGC is 
                                  based on leading log $k_t$-factorization formalism and the b-CGC saturation model. In the right and the left panel we show the results with mini-jet mass $m_{jet}=30$ MeV and  $m_{jet}=1, 5, 10$ MeV respectively. The theoretical uncertainties of about $5\%$ due to fixing the over-all normalization at RHIC are not shown. The experimental data are from the ALICE (preliminary) \cite{alice1}, PHOBOS and BRAHMS collaborations \cite{pa-rhic}.  }
\label{f-alice1}
\end{figure}

After we submitted our paper to archive, the ALICE collaboration released the first p+Pb data from a short pilot run performed in preparation for the full p+Pb physics run scheduled for the beginning of 2013. Here, we compare some of our predictions already presented in this paper with the ALICE preliminary data \cite{alice1,alice2}.  In \fig{f-alice1}, we show  our results for the charged-particle pseudorapidity density in non-single diffractive p+Pb collisions at $\sqrt{S}=5.02$ TeV.  In order to compare with ALICE data,  we have also accounted for the boost of the $\eta=0$ lab frame by adding a rapidity shift of $\Delta y=-0.465$. The details of calculation can be found in Secs. II, III.  We recall that the main theoretical uncertainties in our approach is due to our freedom to choose the value of mini-jet mass $m_{jet}$ appeared in \eq{PO1}. Unfortunately, RHIC data alone is not enough to uniquely fix the value of $m_{jet}$. In \fig{f-alice1}, we show our results with different values of $m_{jet}$. All assumed values of mini-jet within $m_{jet}=0.001\div 0.03$ GeV give a good description of RHIC data for charged hadron multiplicity.  In \fig{f-alice1} (right panel), we show our prediction with $m_{jet}=0.03$ GeV which was available before the ALICE data and it is in agreement with data within about  (or less than) $10\%$ errors. However, it appears that $m_{jet}\approx 5$ MeV gives the best description of the ALICE data with less than $4\%$ errors (see \fig{f-alice1} left panel). This value is remarkably similar to current quark masses. In \fig{fig-m}, we also assumed a fixed $m_{jet}= 5$ MeV for all different centralities.
 We recall that using the $k_T$ factorization, one needs to rewrite the rapidity $y$ distribution in terms of pseudorapidity via $y(h)=\frac{1}{2}\log\frac{\sqrt{\cosh^2 \eta+\mu^2}+\sinh \eta}{\sqrt{\cosh^2 \eta+\mu^2}-\sinh \eta}$ with the Jacobian of rapidity-pseudorapidity transformation $h=\partial y/\partial \eta$. The scale $\mu$ is determined by the typical mini-jet mass and its transverse momentum.  Note that different CGC calculations based on the $k_T$ factorization but using different saturation models (KLN \cite{kln}, IP-Sat \cite{raju-rpa}, rcBK \cite{j2} and b-CGC in this paper) have employed different definition for the scale $\mu$: 
\bea
\mu^2&=&\frac{m_{jet}^2}{P^2}=\frac{0.24^2-0.0035\,\eta\,[N_{part(Pb)}-1]}{(0.13+0.32(\sqrt{S}/1\text{TeV})^{0.115})^2}. \hspace{3.9cm} \text{Ref.\,\cite{kln}: KLN}. \nonumber\\
\mu^2&=&\frac{m_{jet}^2}{p_T^2},\,\,\,\,\, \text{with}\,\,\, m_{jet}=0.2\div 0.4\,\text{GeV}. \hspace{4.6cm} \text{Ref.\,\cite{raju-rpa}: IP-Sat}. \nonumber\\
\mu^2&=&\frac{m_{jet}^2}{P^2}= \frac{0.35^2}{(0.13+0.32(\sqrt{S}/1\text{TeV})^{0.115})^2}.  \hspace{4cm} \text{Ref.\,\cite{j2}: rcBK}. \nonumber\\
\mu^2&=&\frac{m_{jet}^2}{p_T^2},\,\,\,\,\, \text{with}\,\,\, m_{jet}=m_{\text{current quark}}\approx 0.001\div 0.01\,\text{GeV}. \hspace{1.6cm} \text{This paper: b-CGC}. \nonumber\
\eea
 Unfortunately, the value of $m_{jet}$ is interconnected with both soft and hard physics and its true value cannot be determined at the current approximation. We recall also that the value of mini-jet mass  changes the over-all prefactor behind the $k_T$ factorization indicating that $m_{jet}$ may mimic some higher order corrections. Nevertheless, our freedom to chose different values for $\mu$ or $m_{jet}$ may bring uncertainties as large as $5\div 15\%$ at the LHC. Note also that currently all the CGC predictions for charged hadron multiplicity have been based on fixed-coupling $k_T$-factorization formula while incorporating the running-coupling phenomenologically.  However, recently, running coupling corrections for the lowest-order gluon production in high energy hadronic and nuclear scatterings was calculated \cite{ktf} and it was 
 conjectured how running coupling corrections may enter the full fixed-coupling $k_T$-factorization formula. This formulation is yet to be explored for future phenomenological applications.

\begin{figure}[t]                                                            
                                \includegraphics[width=7 cm] {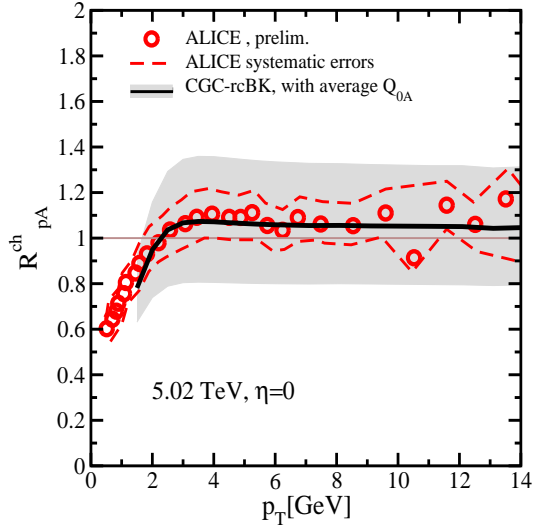}                               
\caption{The nuclear modification factor $R_{pA}^{ch}$ for inclusive charged hadrons production in minimum-bias p+Pb collisions at $\sqrt{S}=5.02$ TeV at $\eta=0$
               obtained  with the solutions of the rcBK with average initial saturation scale for nucleus  $Q_{0A}^2=N\,0.168\,\text{GeV}^2$ with $N=4\div 6$ (grey area) and  $N=5$ (black curve) constrained in \eq{qa}. The theory curves are taken from \fig{rp-h}. The preliminary (prelim.) experimental data are from the ALICE collaboration \cite{alice2}. The dashed lines correspond to maximum systematics errors of experimental data points. }
\label{f-alice2}
\end{figure}
Finally in \fig{f-alice2}, we compare our predictions shown in \fig{rp-h} with preliminary ALICE data for the nuclear modification factor $R_{pA}^{ch}$ of inclusive charged hadrons production in minimum-bias p+Pb collisions at $\sqrt{S}=5.02$ TeV at $\eta=0$. The ALICE data are in excellent agreement with our predictions (shown in \fig{rp-h}) with the solutions of the rcBK evolution equation with average initial saturation scale for nucleus  $Q_{0A}^2=N\,0.168\,\text{GeV}^2$ with $N\approx 5$ constrained in \eq{qa}.  It is remarkable that the preferred value of $N$ corresponds to the average value of $Q_{0A}$ extracted from other reactions given in \eq{qa}. Unfortunately, ALICE data points have rather large systematics errors, nevertheless, it is seen that ALICE data already put a strong extra constrain in \eq{qa} and prefers value within $N\approx 4\div 6$ with $\alpha_s^{in}\approx 0$ (shown in \fig{rp-h} with grey area). Although, a bigger $N$ with larger  $\alpha_s^{in}$  also cannot be ruled out at the moment. As we already stressed the scale of $\alpha_s^{in}$ cannot be determined with the current approximation and it requires a full NNLO calculation which is not yet available. Therefore, our freedom in choosing the value of $\alpha_s^{in}$ in hybrid factorization formula \eq{qa} brings rather large uncertainties.  One of the most remarkable feature of the ALICE data for $R_{pA}^{ch}$ is the fact that data seem to rule out any (or strong) Cronin-type peak. Although, the experimental error bars are still rather large to make any firm conclusion, but if this feature of data remains intact with more precise experimental data, this can be considered as another important evidence in favor of the CGC approach as this feature was already predicted, see discussion in Sec III. The upcoming measurement of $R_{pA}^{ch}$ at forward rapidities at the LHC can be considered as a crucial further test of the CGC approach and provide undoubtedly valuable information about the saturation dynamics and the true value of  the initial saturation scale of heavy nucleus $Q_{0A}$.

\end{document}